\documentclass[aps, twocolumn, nofootinbib,preprintnumbers,eqsecnum,superscriptaddress]{revtex4-1}

\usepackage[usenames, dvipsnames]{color}

\normalsize

\newcommand{\nb}[1]{\color{blue}}

\newcommand{\HL}[1]{{\textcolor{magenta}{#1}}}

\newcommand{\hl}[1]{\color{magenta}}

\usepackage[
	 pagebackref=false,
	 colorlinks=true,
  linkcolor=blue,
  urlcolor=blue,
  filecolor=black,
  citecolor=blue,
  pdfstartview=FitV,
  pdftitle={},
  pdfauthor={},
  pdfsubject={},
  pdfkeywords={},
  pdfpagemode=None,
  bookmarksopen=true
  ]{hyperref}

\usepackage[normalem]{ulem}
\usepackage{amsmath}
\usepackage{enumerate}
\usepackage{amsfonts}
\usepackage{epsfig}
\usepackage{mathbbol}

\def\Tr{\mathop{\rm Tr}}

\newcommand\half{{\ensuremath{\frac{1}{2}}}}

\newcommand\field[1]{{\ensuremath{\mathbb{{#1}}}}}

\newcommand\vev[1]{{\ensuremath{\left\langle{#1}\right\rangle}}}

\newcommand\ket[1]{\ensuremath{\lvert{#1}\rangle}}
\newcommand\bra[1]{\ensuremath{\langle{#1}\rvert}}

\newcommand{\RR}{\field{R}}

\newcommand{\be}{\begin{equation}}
\newcommand{\ee}{\end{equation}}
\newcommand{\bea}{\begin{eqnarray}}
\newcommand{\eea}{\end{eqnarray}}
\newcommand{\bega}{\begin{gather}}
\newcommand{\eega}{\end{gather}}

\newcommand{\bi}{\begin{itemize}}
\newcommand{\ei}{\end{itemize}}
\newcommand{\ben}{\begin{enumerate}}
\newcommand{\een}{\end{enumerate}}
\newcommand{\bca}{\begin{cases}}
\newcommand{\eca}{\end{cases}}
\newcommand{\bln}{\begin{align}}
\newcommand{\eln}{\end{align}}
\newcommand{\bst}{\begin{split}}
\newcommand{\est}{\end{split}}
\def\ie{\begin{equation}\begin{aligned}}
\def\fe{\end{aligned}\end{equation}}
\newcommand{\bma}{\le(\begin{matrix}}
\newcommand{\ema}{\end{matrix}\ri)}

\newcommand\al{{\alpha}}
\def\b{{\beta}}

\newcommand\sig{\sigma}

\newcommand\lam{\lambda}

\newcommand\da{{\dagger}}

\newcommand\ov{\over}
\newcommand\ha{{\half}}

\def\le{\left}
\def\ri{\right}

\newcommand\sE{{\ensuremath{{\mathcal E}}}}

\newcommand\sI{{\ensuremath{{\mathcal I}}}}

\newcommand\sH{{\ensuremath{{\mathcal H}}}}

\newcommand\sL{{\ensuremath{{\mathcal L}}}}

\newcommand\sO{{\ensuremath{{\mathcal O}}}}

\newcommand\sR{{\mathcal R}}
\newcommand\sS{{\mathcal S}}

\newcommand\sZ{{\mathcal Z}}

\newcommand{\Zpt}{\sZ_n^{(\text{PT})}}

\begin{document}

\title{Bound entanglement in thermalized states and black hole radiation}

\preprint{MIT-CTP/5332}

\author{Shreya Vardhan}
\thanks{These authors contributed equally and significantly to the work.}
\affiliation{Center for Theoretical Physics, 
Massachusetts Institute of Technology, Cambridge, MA 02139 }

\author{Jonah Kudler-Flam}
\thanks{These authors contributed equally and significantly to the work.}
\affiliation{Kadanoff Center for Theoretical Physics, 
University of Chicago, Chicago, IL 60637 }

\author{Hassan Shapourian}
\affiliation{Microsoft Station Q, 
Santa Barbara, CA 93109
 }

\author{Hong Liu}
\affiliation{Center for Theoretical Physics, 
Massachusetts Institute of Technology, Cambridge, MA 02139 }

\begin{abstract}

We study the mixed-state entanglement structure of chaotic quantum many-body systems at late times using the recently developed \textit{equilibrium approximation}. A rich entanglement phase diagram emerges when we generalize this technique to evaluate the logarithmic negativity for various universality classes of macroscopically thermalized states. Unlike in the infinite temperature case, when we impose energy constraints at  finite temperature, the phase diagrams for the logarithmic negativity and the mutual information become distinct. In particular, we identify a regime where the negativity is extensive but the mutual information is sub-extensive, indicating a large amount of \textit{bound entanglement}. When applied to evaporating black holes, these results imply that there is quantum entanglement within the Hawking radiation long before the Page time, although this entanglement may not be distillable into EPR pairs.

 \noindent 
 
\end{abstract}


\maketitle


\bigskip
\bigskip

\section{Motivation and introduction}


The entanglement of a bipartite system in a pure state can heuristically be captured by some number of EPR pairs, as it is always possible to convert the pure state into these EPR pairs and vice versa using local operations and classical communication (LOCC). 
This kind of interpretation becomes more complicated for mixed states. 
The interconversion between mixed states and EPR pairs is in general irreversible; the number of EPR pairs needed to prepare the state using LOCC operations (entanglement cost) can be greater than the number that can be extracted from it (distillable entanglement) \cite{bennett1996mixed}. In particular, there exist bound-entangled states \cite{horodecki1998mixed}, which have non-zero entanglement cost, but from which no EPR pairs can be distilled.

While mixed-state entanglement carries important physical information, 
the corresponding operational measures such as entanglement cost and distillable entanglement
are extremely hard to calculate even for few-qubit systems. The logarithmic negativity provides a more calculable measure \cite{1996PhLA..223....1H,1998PhRvA..58..883Z, 1996PhRvL..77.1413P, eisert1999comparison, 2000PhRvL..84.2726S, vidal2002computable, plenio2005logarithmic}, but is still very difficult to compute in many-body systems and field theories. 

In this paper, we generalize a method developed in~\cite{liu2021entanglement}, called the equilibrium approximation, to obtain the logarithmic negativity of a macroscopically equilibrated mixed state. The approximation applies to general chaotic systems at late times, when the macroscopic properties of the state are close to those of a thermal ensemble,\footnote{Such macroscopic properties include expectation values of local operators.} although the state is far from the thermal density matrix by measures like trace distance.
 We find that there can be a rich entanglement structure, captured by an intricate entanglement phase diagram. A particularly surprising result is that in the thermodynamic limit, there can be a finite region in the parameter space where the logarithmic negativity is extensive, but the mutual information is sub-extensive, implying that there is a large amount of bound entanglement. This phenomenon does not take place in the infinite temperature case previously studied in \cite{2021PRXQ....2c0347S}, but arises in a variety of universality classes of equilibrated states at finite temperature studied here. 

A main physical application we have in mind is 
an evaporating black hole. 
Consider a black hole formed from the gravitational collapse of matter in a pure state. The black hole emits Hawking radiation, and eventually evaporates completely. The dynamics of black holes are expected to be highly chaotic. Hence, if the evaporation process respects the usual rules of quantum mechanics, general results on the quantum-informational properties of a chaotic quantum many-body system can be used to make predictions about the black hole and its radiation. 
The verification of such predictions using gravity calculations can then provide highly nontrivial checks of the consistency of black hole physics with quantum-mechanical principles, and can also lead to new insights into quantum gravitational dynamics. One good example is the celebrated Page curve, a prediction for the time-evolution of the entanglement entropy of the radiation~\cite{page1993average, page1993information}. This curve was recently derived in \cite{penington2020entanglement, Almheiri:2019psf, penington2019replica, almheiri2020replica} with gravity calculations, which not only confirmed that black holes obey unitarity, but also revealed new geometric features known as ``islands" and ``replica wormholes." 

\begin{figure}[] 
\centering 
\includegraphics[width=.4\textwidth]{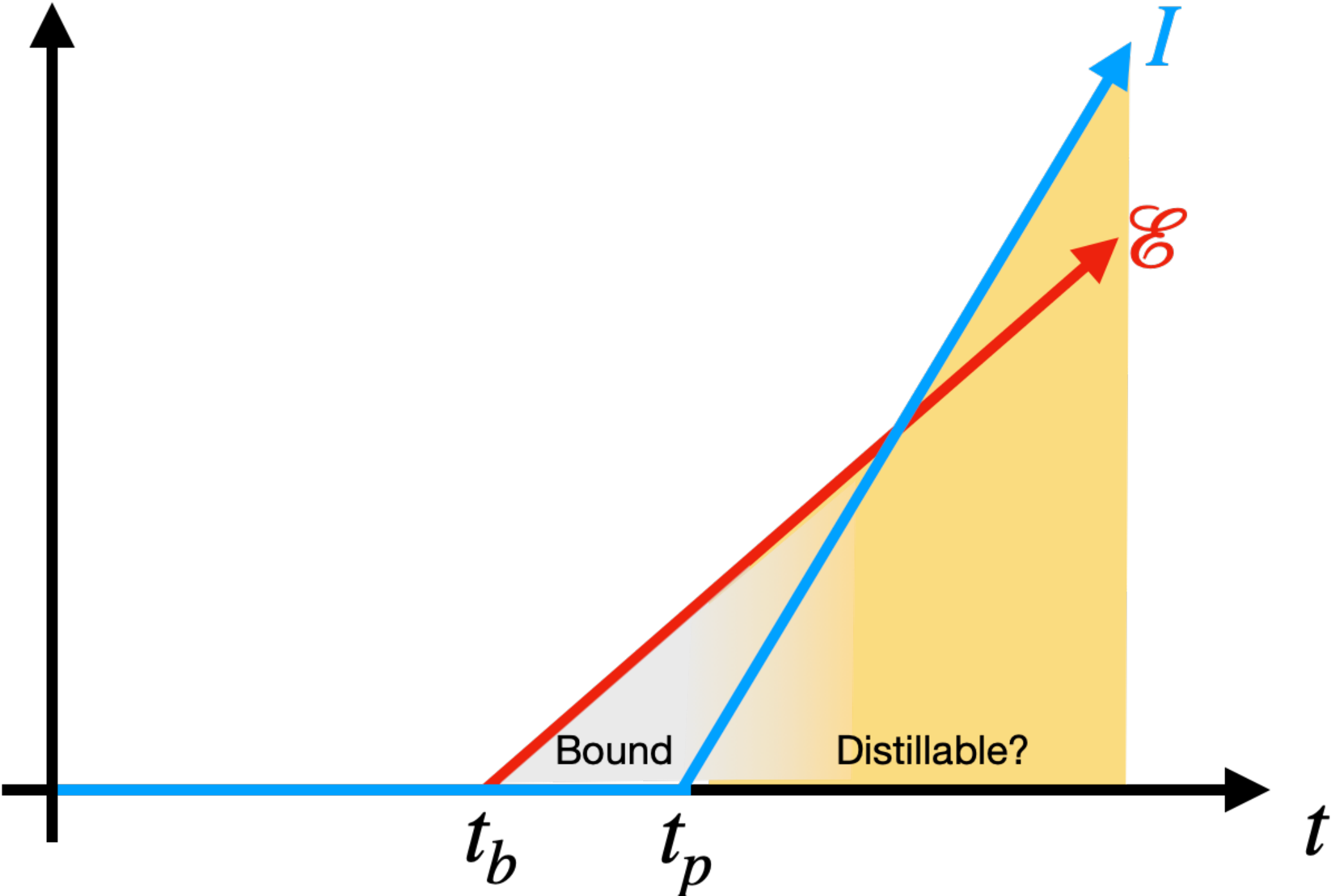}
\caption{The general behavior of logarithmic negativity and mutual information is shown for finite temperature equilibrated pure states, specifically evaporating black holes. While the mutual information within the radiation does not become extensive until the Page time $t_p$, the negativity becomes large at the earlier time $t_b$, signaling the existence of quantum correlations in the radiation prior to the Page time that cannot be distilled into EPR pairs. After the Page time, we expect the entanglement to be distillable but do not have a rigorous proof.}
\label{fig:bound_cartoon}
\end{figure}

The evaporation process is very slow in microscopic scales, so that at any time in the process, we can treat the remaining black hole as well as the radiation as being in macroscopic equilibrium,\footnote{The two systems can be in separate macroscopic equilibria, because in general the radiation is separated from the black hole after it is emitted.} even though the full system is in a pure state, and the reduced density operator of each subsystem can also be far from the density operators of thermal ensembles. 
Page considered the von Neumann entropies of the black hole and the radiation at infinite temperature. 
However, in general it is important to study the entanglement structure at finite temperature, and also to probe it with other quantum-informational quantities such as logarithmic negativity. 
The general results obtained in this paper serve as predictions for the entanglement structure {\it within} the radiation\footnote{These results also provide hints of multipartite entanglement among the black hole and different parts of the radiation.} at both infinite and finite temperatures. At infinite temperature, before the Page time, the radiation is maximally entangled with the black hole, and there is no entanglement within the radiation itself. In contrast, at finite temperature we find that there is a new time scale $t_b$ {\it before} the Page time $t_p$, when nontrivial entanglement within the radiation starts to emerge (see Fig.~\ref{fig:bound_cartoon}). For $t \in (t_b, t_p)$, the entanglement is bound entanglement, that is, it cannot be distilled into EPR pairs using LOCC.\footnote{It remains a logical possibility that EPR pairs may be distillable using positive partial transpose~(PPT) operations.} In this paper, we outline the general ideas and the main results, leaving technical details and further elaboration to~\cite{long_paper}. 

For a recent study of the logarithmic negativity in thermalized chaotic quantum-many body systems, see \cite{2020PhRvB.102w5110L}. For previous studies in gravitational systems, see \cite{2019PhRvD..99j6014K,2019PhRvL.123m1603K,2021JHEP...06..024D,2021arXiv210902649K}. 

\section{Setup} 

Consider a system $A$ in a mixed state $\rho_A$, which is in a macroscopic equilibrium but can be far in trace distance from the usual equilibrium density operators describing thermal ensembles. 
To explore the bipartite entanglement structure of $A$, we would like to evaluate the logarithmic negativity $\sE (A_1, A_2)$ and mutual information $I (A_1, A_2)$ 
between a subsystem $A_1$ and its complement $A_2$ in $A$. The logarithmic negativity $\sE$ is nonzero only if 
$\rho_A$ is not separable,
and can be used to lower-bound the PPT entanglement cost, $\mathcal{E} (A_1, A_2) \leq E_c^{\rm (ppt, exact)}(A_1, A_2)$ \cite{audenaert2003entanglement}.
The mutual information $I$ also contains classical information, but is nevertheless of importance as it upper-bounds the distillable entanglement, $E_d (A_1, A_2) \leq \ha I (A_1, A_2)$ \cite{2004JMP....45..829C}. We will also discuss the behavior of the R\'enyi mutual informations $I_n(A_1, A_2)$ in some cases below. See Appendix \ref{app:info} for the definitions of these different information-theoretic quantities. 

We can imagine that $A$ is embedded in a larger system $S = A\cup B$, with the total system $S$ in a pure state $\ket{\Psi}$ in macroscopic equilibrium\footnote{Note that $A$ and $B$ in principle do not have to be in equilibrium with each other.} and $\rho_A = {\rm Tr}_B \ket{\Psi} \bra{\Psi}$. 
In many situations of interest, such a $B$ naturally exists. For example, for an evaporating black hole, when $A$ is taken to be the Hawking radiation, $B$ is the remaining black hole. 

In~\cite{liu2021entanglement}, it was shown that the quantum-informational properties for a system in such an equilibrated pure state $\ket{\Psi}$ can be calculated from properties of an equilibrium density matrix
\be \label{heno}
\rho^{\rm (eq)} := {1 \ov Z_1} \sI_\al, \qquad Z_1:= \Tr \sI_\al ,
\ee
which has the same macroscopic thermodynamic behavior as $\ket{\Psi}$. Here $\al$ denotes macroscopic equilibrium parameters such as temperature or chemical potential. Specification of $\sI_\al$ can be viewed as specifying the universality class of an equilibrated pure state.

From~\cite{liu2021entanglement}, the R\'enyi partition function for a subsystem $R$ can be obtained by 
\be\label{fen}
\sZ_{n, R} := \, \text{Tr} \rho_R^n \approx \, {1 \ov Z_1^n} \sum_{\tau \in \sS_n} \vev{\eta_R \otimes e_{\bar R} | \sI_\al , \tau} ,
\ee
where $\tau$ is an element of the permutation group $\sS_n$, and $\bar R$ is the complement of $R$ in $S$. See Appendix~\ref{app:eq} for other notations in this expression. When~\eqref{fen} can be expressed as an analytic function of $n$, the von Neumann entropy for $R$ can be obtained by analytic continuation of \eqref{fen}. Alternatively, we can use~\eqref{fen} to calculate the resolvent, 
\be 
\sR(\lambda) :=\text{Tr}\left(\frac{1}{\lambda \mathbf{1} -\rho_R}\right) = \frac{1}{\lambda}\sum_{n=0}^{\infty} \frac{1}{\lambda^{n}} \sZ_{n, R} \ , 
\label{Rdef}
\ee 
which can be used to find the spectral density of $\rho_R$, 
\be 
D (\lam) = \frac{1}{\pi } ~\lim_{\epsilon \rightarrow 0}~ {\rm Im} \, \sR(\lambda- i\epsilon), \quad \lam \in \RR \ \, , 
\ee
and hence the von Neumann entropy 
\be 
S(R) = -\int d \lam \, D(\lam) \, \lam \log \lam \, . 
\label{Sdef}
\ee

Equations~\eqref{fen}--\eqref{Sdef} can be used to calculate the mutual information $I(A_1, A_2)$ by taking $R$ to be the different subsystems $A$, $A_1$, $A_2$.

The methods of~\cite{liu2021entanglement} can also be generalized to calculate the partial transpose partition function
\begin{align}\label{repa}
 \mathcal{Z}_n^{(\rm PT)} &:= {\rm Tr}_A \left( \rho_{A}^{T_2} \right)^n \nonumber
 \\ &\approx \frac{1}{Z_1^n} \sum_{ \tau}\bra{\eta_{A_1} \otimes \eta^{-1}_{A_2} \otimes e_{B}}\mathcal{I}_{\alpha},\tau\rangle,
\end{align}
where $T_2$ denotes partial transpose of $\rho_A$ with respect to $A_2$. 
When~\eqref{repa} is analytic in \emph{even} $n$, the logarithmic negativity can be found from analytic continuation as $\mathcal{E}(A_1, A_2) = \lim_{n \to \ha} \log \mathcal{Z}_{2n}^{(\rm PT)}$. Alternatively, it can be calculated from the resolvent for $\rho_A^{T_2}$ as
\bega \label{Rndef}
R_N(\lambda) :=\text{Tr}\left(\frac{1}{\lambda \mathbf{1} -\rho_A^{T_2}}\right) = \frac{1}{\lambda}\sum_{n=0}^{\infty} \frac{1}{\lambda^{n}} \sZ_n^{\rm (PT)}, \\
D_N (\lam) = \frac{1}{\pi } ~\lim_{\epsilon \rightarrow 0}~{\rm Im} \, R_N (\lambda- i\epsilon), \quad \lam \in \RR \ , \\
\sE(A_1, A_2) = \log\left(\int d \lam \, D_N(\lam) \, |\lambda|\right) \, . 
\label{erdef}
\end{gather} 
where $D_N({\lam})$ is the spectral density of $\rho_A^{T_2}$.

We will consider $\sE(A_1, A_2)$ and $I(A_1,A_2)$ at leading order in the thermodynamic limit. In this limit, \eqref{fen} and \eqref{repa} can both be approximated by terms from a subset of permutations $\tau$, which give the dominant contribution. These sets of permutations can change as we vary two parameters
\be \label{lamd}
\lam := {S_{A_1}^{(\rm eq)} \ov S_A^{(\rm eq)}} , \qquad c := {S_A^{(\rm eq)} \ov S_A^{(\rm eq)} + S_B^{(\rm eq)}},
\ee
where $S_{A_1, A, B}^{\rm (eq)}$ are respectively the von Neumann entropies for $A_1, A, B$ in the state $\rho^{(\rm eq)}$. These parameters can be seen as a way of measuring the relative sizes of the subsystems in the general case where the system $S = A \cup B$ is inhomogeneous; when the full system is homogeneous, $\lam$ and $c$ 
are simply the volume fractions of various subsystems, $\lam = {V_{A_1}/ V_A}$ and $c = {V_A /( V_A + V_B)}$, where $V_{A_1}, V_A, V_B$ are respectively
the volumes of $A_1, A, B$. The change in the dominant contribution on varying $c$ and $\lambda$ leads to qualitative changes in the behavior of $\sZ_n^{\rm (PT)}$ and $\sZ_{n, R}$, and correspondingly of $\sE(A_1, A_2)$ and $I(A_1,A_2)$. We refer to such changes as entanglement phase transitions.

\section{Entanglement phase diagram at infinite temperature} 
\label{sec:inf_temp}

\begin{figure}[] 
\centering 
\includegraphics[width=8cm]{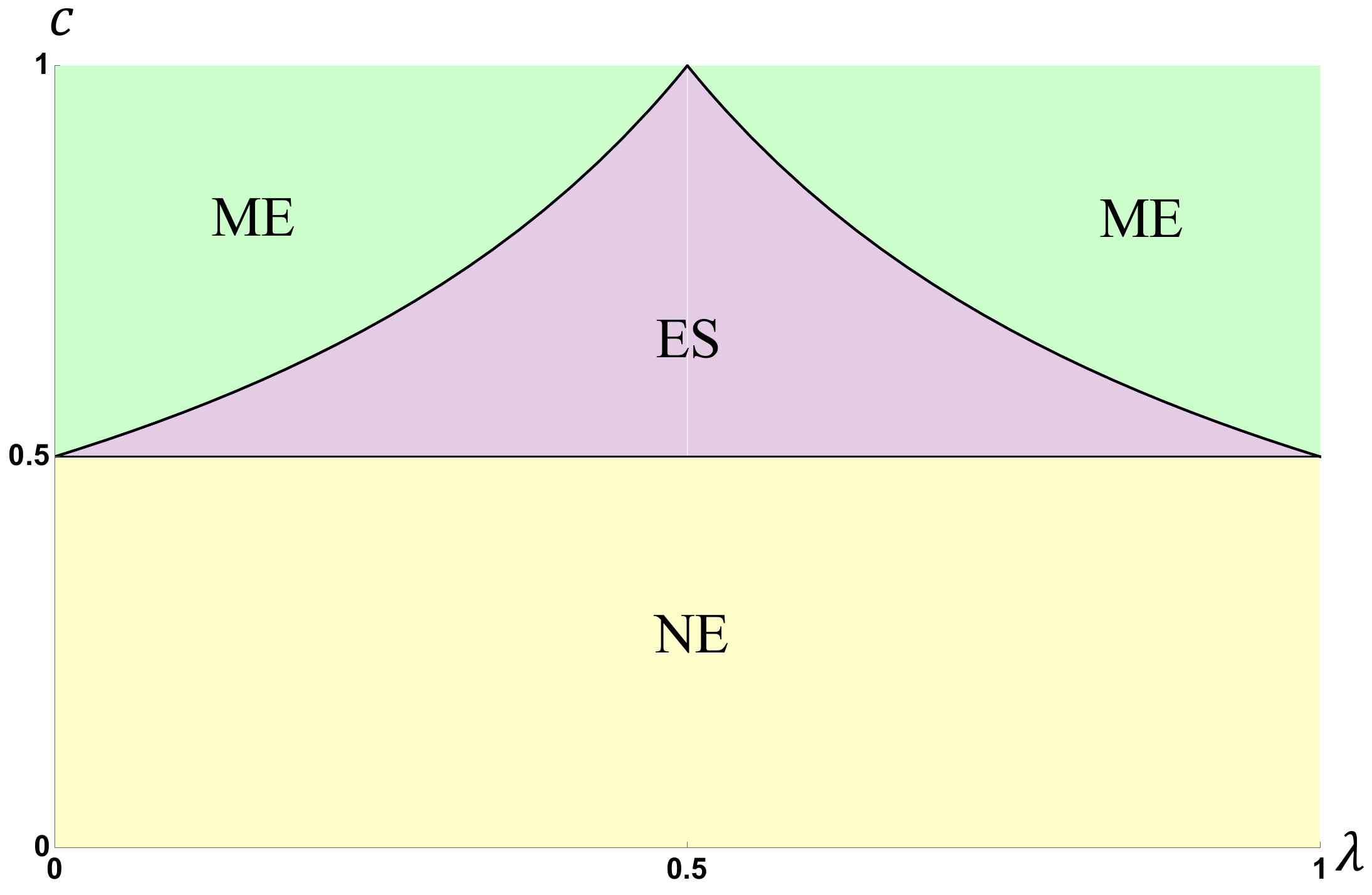}
\caption{Entanglement phase diagram for the infinite temperature equilibrated pure state.}
\label{fig:inf_phases}
\end{figure}

For an equilibrated pure state at infinite temperature, $\sI_\al$ is given by the identity operator ${\mathbf 1}_{AB}$, and 
we find a universal entanglement phase structure.
The partial transpose partition function $\Zpt$, logarithmic negativity $\sE$, R\'enyi mutual information $I_n$, and mutual information $I$ all have the same $n$-independent phase structure. The structure also coincides with that obtained from the Haar average of a random state~\cite{2021PRXQ....2c0347S}. The phase diagram is given in Fig.~\ref{fig:inf_phases} and can be summarized as follows: 

\vspace{0.3cm}

\noindent {\bf 1. Phase of no entanglement (NE)}.\footnote{``No entanglement'' here should be understood as no ``volume-law'' entanglement, i.e.~there is no contribution at the order of $O(\log d_A)$.} \\For $c<\ha$, or equivalently $S^{\rm (eq)}_A < S^{\rm (eq)}_B$, we find 
\be \label{ehn}
\sE(A_1, A_2) =I (A_1, A_2) = 0 
\ee
and furthermore all the R\'enyi mutual informations vanish. These results imply that $\rho_A$ is close to being maximally mixed, and all degrees of freedom of $A$ are maximally entangled with those in $B$.



\vspace{0.3cm}

\noindent \textbf{2. Maximally entangled phase (ME)}.\\ For $c>\ha, \, \lambda>\frac{1}{2c}$, or equivalently $S_B^{(\rm eq)} < S_A^{(\rm eq)},\; S_{A_2}^{(\rm eq)} < \ha (S_A^{(\rm eq)} - S_B^{(\rm eq)})$, we find 
\begin{gather} 
\sE(A_1, A_2) = \ha I (A_1, A_2) = S_{A_2}^{(\rm eq)},
\label{legn}
\end{gather} 
which is the maximal 
value that $\sE (A_1, A_2)$ and $I(A_1, A_2)$ can have, and implies that $A_2$ is maximally entangled with $A_1$. In this regime, we also have $S_{A_2}^{(\rm eq)} + S_B^{(\rm eq)} < \ha (S_A^{(\rm eq)} + S_B^{(\rm eq)}) < S_{A_1}^{(\rm eq)}$,  i.e.~the effective number of degrees of freedom in $A_2$ and $B$ together is smaller than that in $A_1$. Thus both $A_2$ and $B$ should be maximally entangled with $A_1$. 
Similarly for $c>\ha, ~\lambda<1-\frac{1}{2c}$, 
$A_1$ is maximally entangled with $A_2$, with the $\sE$ and $I$ obtained by exchanging $A_1$ and $A_2$ in
\eqref{legn}. 

\vspace{0.3cm}

\noindent \textbf{3. Entanglement saturation phase (ES)}. \\ For $c>\ha,\; 1-\frac{1}{2c}< \lambda< \frac{1}{2c}$, or equivalently $S_B^{(\rm eq)} < S_A^{(\rm eq)},\; \ha (S_A^{(\rm eq)} - S_B^{(\rm eq)}) < S_{A_1}^{(\rm eq)} < \ha (S_A^{(\rm eq)} + S_B^{(\rm eq)})$, we have
\bega 
\sE(A_1, A_2) = \ha I (A_1, A_2) =
\ha (S_A^{(\rm eq)} - S_B^{(\rm eq)}) \ .
\label{neg_inf_es}
\end{gather} 
Both the negativity and the mutual information depend only on the difference $S_A^{(\rm eq)} - S_B^{(\rm eq)}$, and do not change as we vary the size of $A_1$ (as long as we stay in the aforementioned parameter range). Note that in both \eqref{legn} and \eqref{neg_inf_es}, $\sE$ is half of the corresponding values for $I_{1/2}$ like in a pure state, even though $\rho$ is mixed. 
 There is a simple intuitive interpretation of~\eqref{neg_inf_es} in terms of bipartite entanglement among pairs of subsystems: since $S_B^{(\rm eq)} < S_A^{(\rm eq)}$, $\log d_B = S_B^{(\rm eq)} $ degrees of freedom in $A$ are entangled with $B$, and the remaining $\log d_A - \log d_B = S_A^{(\rm eq)} - S_B^{(\rm eq)} $ are entangled between $A_1$ and $A_2$. We should emphasize, however, that this ``mechanical'' way of assigning entanglement likely does not 
reflect the genuine entanglement structure of the system in this phase, and there are indications of significant multi-partite entanglement in this phase~\cite{long_paper}.

\section{The equilibrium approximation for mixed-state entanglement at finite temperature}

The infinite temperature case only applies to a system with a finite-dimensional Hilbert space at sufficiently high energies or without energy conservation. When the energy of a system is not large enough, or if a system has an infinite-dimensional Hilbert space such as in a field theory, energy constraints must be imposed. 
Now there are many more possibilities for $\sI_\al$, which depend on the ensemble we choose. 

 At finite temperature, in general each of the infinite number of quantities $\Zpt, I_n$ 
 gives rise to a different phase diagram, revealing intricate patterns of entanglement structure. We will focus on the behavior of the logarithmic negativity $\sE$ and the mutual information $I$. 
A significant technical complication at finite temperature is that the extraction of the logarithmic negativity and the von Neumann entropies using analytic continuation becomes a priori unreliable near phase boundaries, due to the non-uniform dependence on $n$ of $\Zpt$ and the R\'enyi entropies. $\sE$ and $I$ must be calculated using the resolvents in \eqref{Rdef} and \eqref{Rndef}, which fortunately may be done (in Appendix \ref{app:resolvent}) for some choices of $\sI_\al$ and used to illustrate the general structure. 

Consider first the mutual information. At finite temperature, to leading order in volume, the equilibrium approximation for $\sZ_{n, R}$ leads to the following approximation for the von Neumann entropy
\be 
S_{R}= {\rm min} \le(S_{R}^{({\rm eq})} , S_{\overline{R}}^{({\rm eq})}\ri)\, . \label{page_gen}
\ee
In~\cite{liu2021entanglement}, this finite-temperature generalization of Page's formula was argued on the basis of analytic continuation, which is only reliable in cases where $R \cup \overline{R}$ is a homogeneous system and the Page transition point in $\sZ_{n,R}$ is independent of $n$. 
However, this statement remains true 
in all cases of $\sI_\al$ that we have studied using the resolvent, including inhomogeneous cases (see Fig.~\ref{fig:resolvent_checks}~(b)). We will assume that \eqref{page_gen} holds in general. Given that the von Neumann entropy of a thermal density operator is extensive ($S_{A_1}^{( {\rm eq})} + S_{A_2}^{( {\rm eq})} = S_{A}^{({\rm eq})}$), 
we then find that $I(A_1, A_2)$ has exactly the same behavior as at infinite temperature, with the same expressions as those in~\eqref{ehn}-\eqref{neg_inf_es},\footnote{The corresponding $S_A^{(\rm eq)}$ and so on should be replaced by their finite temperature counterparts.} and the same phase diagram given by Fig.~\ref{fig:inf_phases}. 

Now consider the logarithmic negativity $\sE(A_1, A_2)$, which reveals a much richer structure. 
The precise phase diagram and the expressions for $\sE$ in different phases depend on the choice of $\sI_\al$, but in all cases we have studied, there are analogs of each of the infinite temperature phases NE, ES, and ME. In all cases, these come from three different choices of the dominant permutations in \eqref{repa}, which are discussed in Appendix \ref{app:eq}. 
A most surprising feature, which appears to be generic in the examples where the resolvents can be explicitly calculated, is that there is a regime for an $O(1)$ range of the parameter $c$ where $\sE$ is extensive while $I$ is sub-extensive. It is generally believed that since the mutual information contains both quantum and classical correlations, there cannot be any 
 volume-like quantum entanglement when it is 
sub-extensive. Our results indicate that this intuition cannot be correct. We will elaborate further on this point below. Another surprising feature is that in addition to the phases that can be deduced by analytic continuation from the phases of $\Zpt$, there can be new phases in $\sE$ which do not correspond to any of the phases of $\Zpt$.
Note that in all cases below, $S^{\rm (eq)}_{n, R}$ refers to the $n^{th}$ R\'enyi entropy of subsystem $R$ in the thermal state $\rho^{\rm (eq)}$ for that case. 

Consider first the canonical ensemble 
\be 
\sI_\al = e^{-\b_A H_A} \otimes e^{-\b_B H_B}\, , \label{can}
\ee
where we have allowed $A$ and $B$ to have different inverse temperatures $\b_A, \b_B$. Using~\eqref{repa}, the following forms of the negativity can be deduced from the phases of $\sZ_n^{\rm (PT)}$ by analytic continuation
\begin{align} 
\sE_{NE} &= 0 \label{fin_zero}\\
 \sE_{ME} &= S^{\rm (eq)}_{\frac{1}{2},\, A_1}(\beta_A)\quad \text{ or } \quad \sE_{ME} = S^{\rm (eq)}_{\frac{1}{2},\, A_2}(\beta_A) \label{fin_me}\\
\sE_{ES} &= \ha \le(S^{\rm (eq)}_{{1 \ov 2}, A}(\beta_A) 
- S_{2, B}^{\rm (eq)} (\beta_B)\ri) \label{fin_es}
\end{align}
These give generalizations of the infinite temperature values in \eqref{ehn}-\eqref{neg_inf_es}. Moreover, comparison of $\sE_{ES}$ and $\sE_{NE}$ would suggest that the transition line between the NE and ES phases is given by the condition 
\be\label{ehv}
S^{\rm (eq)}_{{1 \ov 2}, A} 
= S_{2, B}^{\rm (eq)} \ . 
\ee
Given that $S_n$ monotonically decreases with $n$, we have $S_{\frac{1}{2}, A}^{\rm (eq)} > S_{A}^{\rm (eq)} $
and $S^{\rm (eq)}_{2, B} < S^{\rm (eq)}_{B} $. Hence, the transition~\eqref{ehv} must happen for
$S_{A}^{\rm (eq)} < S^{\rm (eq)}_{B} $,  i.e.~at some $c_0 < \ha$, so that there is a region in the phase diagram where the logarithmic negativity is volume-like, but the mutual information is not yet volume-like. In the case where $A$ is at infinite temperature while $B$ is at finite temperature, we found the exact resolvent numerically in the NE and ES phases, and confirmed \eqref{fin_es} and \eqref{ehv}. See Fig.~\ref{fig:resolvent_checks}~(a). We can also calculate 
$\sE$ analytically from the resolvent in a certain approximation for the NE and ES regimes to confirm \eqref{fin_es} and \eqref{ehv}. 
A special example is the toy model of black hole evaporation in Jackiw-Teitelboim (JT) gravity discussed in~\cite{penington2019replica}, 
where 
\bea 
& S_{2, B}^{(\rm eq)} = S_0 + s_2 (\b) , \quad S_{B}^{(\rm eq)} = S_0 + s_1 (\b) , \nonumber \\ & s_1 (\b) > s_2 (\b) \sim O(1) \label{jt}
\eea
for which we have $c_0 = \ha \le(1 - {s_1 - s_2 \ov 2 S_0} \ri)$. In this model, due to the special structure of the density of states for JT gravity, the difference between $c_0$ and $c$ turns out to be subleading in $S_0$; for higher-dimensional gravity systems, this difference would be $O(1)$. The negativity phase diagram for this model is also discussed in \cite{2021arXiv210902649K,sb_forthcoming}. 

As another example, suppose the total energy in $A$ is conserved and it equilibrates to the microcanonical ensemble, while $B$ equilibrates to infinite temperature. For this case, 
\be \label{mic_case1}
\sI_{\al} = \sum_{E^{A_1}_{a_1} + E^{A_2}_{a_2} \in I_{E, \Delta}} \ket{a_1}\bra{a_1} \otimes \ket{a_2}\bra{a_2} \otimes \mathbf{1}_B ,
\ee
where $E^{A_s}_{a_s}$ refer to energies in $A_s$, and $I_{E, \Delta}$ is a narrow energy interval $[E-\Delta, E+\Delta]$, where $\Delta$ is some $O(1)$ constant. As we vary the volumes of the different subsystems, we fix the average energy density in $A$ to a value $\frac{E}{V_A} = \epsilon$, and the infinite temperature equilibrium entropy for $B$ is $S_{n, B}^{(\rm eq)} = \log d_B$ for all $n$. In this case, from~\eqref{repa} and analytic continuation in $n$ we find 
\begin{align} 
\sE_{NE} &= 0 \label{fin2_zero}\\
 \sE_{ME} &= S^{\rm (eq)}_{\frac{1}{2},\, A_1}\quad \text{ or } \quad \sE_{ME} = S^{\rm (eq)}_{\frac{1}{2},\, A_2} \label{fin2_me}\\
\sE_{ES} &= \ha \le(S^{\rm (eq)}_{{1 \ov 2}, A_1} + S^{\rm (eq)}_{{1 \ov 2}, A_2} 
-\log d_B\ri) \label{fin2_es}
\end{align} 

These expressions resemble~\eqref{fin_zero}-\eqref{fin_es}.\footnote{Note that in microcanonical ensemble in general $S^{\rm (eq)}_{{1 \ov 2}, A_1} + S^{\rm (eq)}_{{1 \ov 2}, A_2} \neq S^{\rm (eq)}_{{1 \ov 2},A} $.} The value $c_0~ < ~\ha$ for the transition between the ES and NE phases is again determined by the condition $\sE_{ES} =0$. 
\eqref{fin2_es} and the transition line from NE to ES can again be confirmed using a resolvent calculation within a certain approximation in this example.

\begin{figure*}[]
\centering
\includegraphics[width = .48\textwidth]{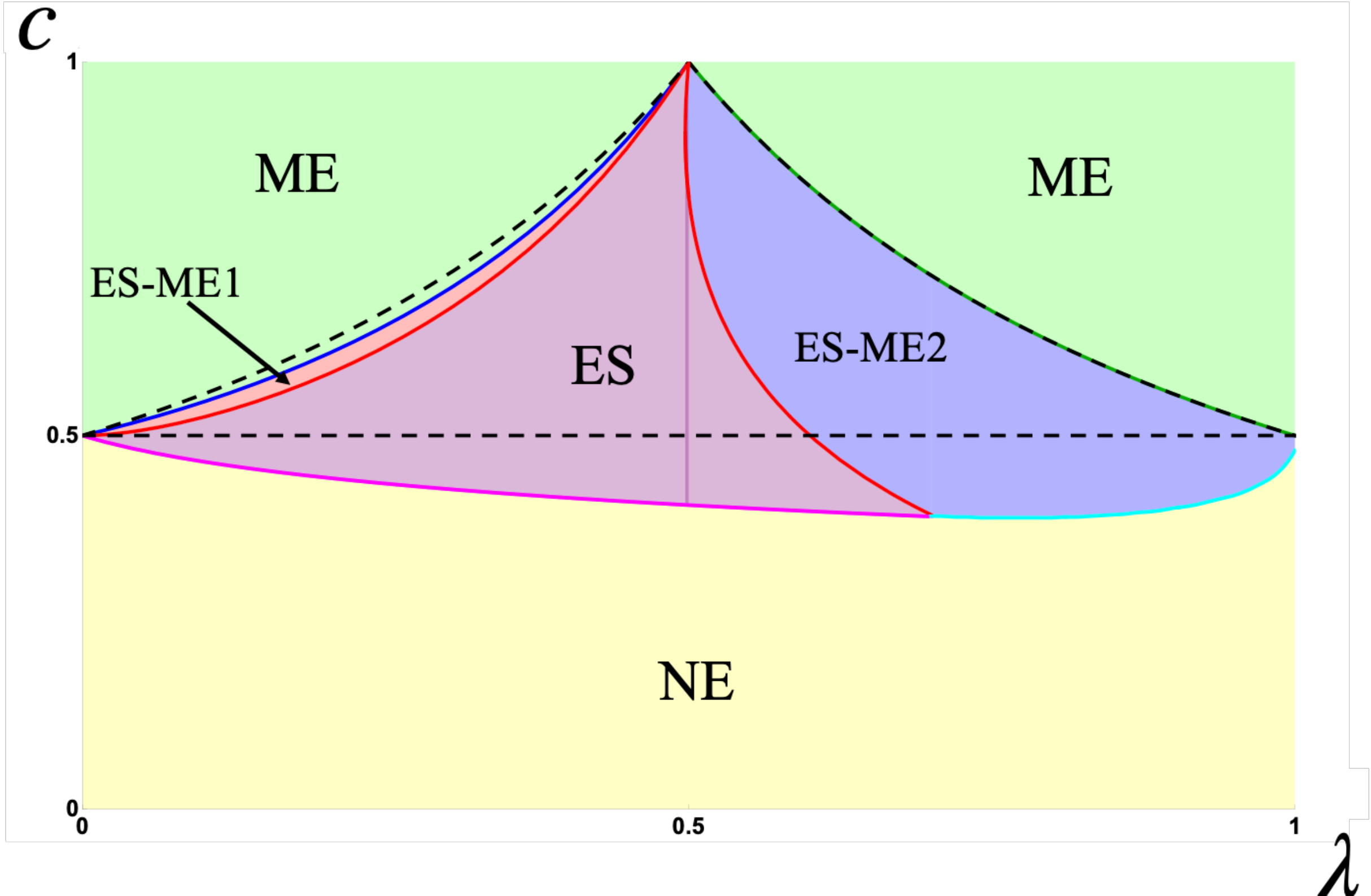}
\includegraphics[width = .48\textwidth]{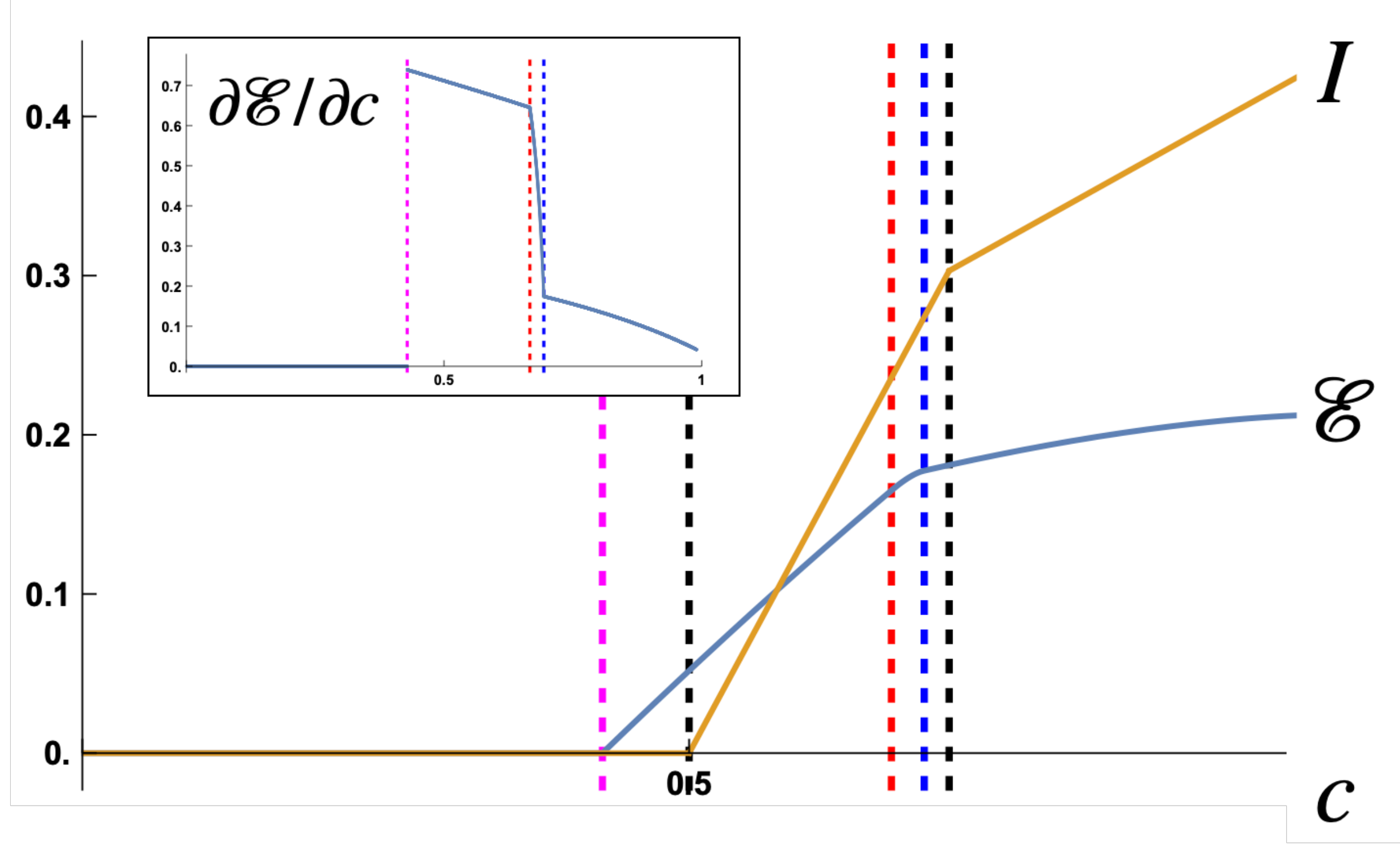}
\caption{Left: the phases of the logarithmic negativity for~\eqref{mic_inf_temp} are shown with the different colored regions. The phase transition lines for the mutual information are shown with the dashed black lines.
The magenta line between NE and ES agrees with setting~$\sE_{ES}$ in \eqref{fin3_es} to zero. Note that the precise shape of the phase transition lines for $\sE$ depends on the average energy density $\epsilon$ and the form of the entropy density $s(\epsilon)$ for the system; here we took $s(\epsilon)= \sqrt{\epsilon}$ and $\epsilon=0.5$. Right: $\sE$ and $I$ along a vertical line at $\lambda=0.3$ in the phase diagram, and the inset shows $\frac{\partial\sE}{\partial c}$ along the same line, with the phase transitions indicated by vertical grid lines. }
\label{fig:phase_diagram_a2_inf}
\end{figure*}

Next, consider an example where the system $A_2$ is taken to be at infinite temperature, while $A_1 B$ is a homogenous system that satisfies energy conservation and equilibrates to the microcanonical ensemble. In this case, 
\be 
\sI_{\alpha} = \mathbf{1}_{A_2} \otimes \sum_{E_p^{A_1} + E_r^B \in I_{E, \Delta}} (\ket{p}\bra{p})_{A_1} \otimes (\ket{r}\bra{r})_{B} .
\label{mic_inf_temp}
\ee
As we vary the volumes of the different subsystems to find the phase diagram, we keep the average energy density in $A_1B$ fixed to some value $\frac{E}{V_{A_1}+V_B}=\epsilon$, and $S_{A_2}^{(\rm eq)} = \log d_{A_2}$. We then find that 
\begin{align} 
\sE_{NE} &= 0 \label{fin3_zero}\\
 \sE_{ME} &= S^{\rm (eq)}_{\frac{1}{2},\, A_1}\quad \text{ or } \quad \sE_{ME} = \log d_{A_2} \label{fin3_me}\\
\sE_{ES} &= \ha \log d_{A_2} + S^{\rm (eq)}_{{1 \ov 3}, A_1} -\ha S^{\rm (eq)}_{A_1} - \ha S_{B}^{\rm (eq)} \label{fin3_es}
\end{align} 
The transition between NE and ES is given by setting $\sE_{ES}$ to zero, and the corresponding $c_0$ is $< \ha$ since $S^{\rm (eq)}_{{1 \ov 3}, A_1} > S^{\rm (eq)}_{A_1}$. These results can again be confirmed from a resolvent calculation. In fact, in this case it is possible to use the resolvent to find the phase diagram for all regimes of $c$ and $\lambda$, as shown in Fig.~\ref{fig:phase_diagram_a2_inf}. In addition to finite temperature generalizations of the NE, ES, and ME phases, where $\sE$ takes the forms \eqref{fin3_zero}-\eqref{fin3_es} expected from analytic continuation, there are also two new phases, which we call ES-ME1 and ES-ME2. These phases cannot be obtained by analytic continuation of $\sZ_n^{\rm (PT)}$, and the expressions for $\sE$ in them cannot be expressed simply in terms of equilibrium R\'enyi entropies. These expressions are given in Appendix \ref{app:resolvent}. While there is a discontinuity in first derivative $\frac{\partial \sE}{\partial c}$ in going from the NE phase to the ES or ES-ME2 phases, for the remaining phase transitions, there is a discontinuity only in the second derivative $\frac{\partial ^2\sE}{\partial c^2}$, as shown in Fig. \ref{fig:phase_diagram_a2_inf}. 

The phase diagram for the mutual information is also shown with dashed lines in Fig. \ref{fig:phase_diagram_a2_inf} for comparison. One feature of the phase diagram is that the logarithmic negativity and the mutual information appear to reflect different physics; in general, at the phase boundaries of the mutual information, there is no qualitative change in the logarithmic negativity and vice versa. The one exception to this in Fig. \ref{fig:phase_diagram_a2_inf} is the transition between the ES and ME phases for the mutual information, and the transition between the ES-ME2 and ME phases for $\sE$, which coincide. This is likely to be a specific feature of this model coming from taking $A_2$ to be at infinite temperature. Also note that while the transition in $I$ at this line is a first-order transition, the transition in $\sE$ is second-order. 

Note that since we only found analytic expressions or detailed numerics for $\sE$ in the earlier examples \eqref{can} and \eqref{mic_case1} in the regime corresponding to the NE and ES phases and the transition between them, there remains the possibility that these examples could also have new phases that cannot be obtained by analytic continuation.

Let us now discuss the operational implications of the above results for $\sE$ and $I$. As mentioned earlier, the distillable entanglement is upper-bounded by half of the mutual information \cite{2004JMP....45..829C}, 
\be
E_d (A_1, A_2) \leq \ha I (A_1, A_2) \label{m_ed}
\ee
while the exact PPT entanglement cost is lower-bounded by the logarithmic negativity \cite{audenaert2003entanglement}, 
\be 
 E_c^{\rm (ppt, exact)}(A_1, A_2) \geq \sE(A_1, A_2)
 \label{e_ec}
\ee 
Hence, in situations where $\sE \gg \frac{1}{2}I$, we can draw the operational conclusion that 
\be 
E_c^{\rm (exact)} \gg E_d^{\rm (exact)}\, . 
\ee
If we assume that $E_c^{\rm (ppt)}$ is also extensive when $E_c^{\rm (ppt, exact)}$ is extensive, then we can make the stronger statement 
\be 
E_c \gg E_d \, . \label{cd}
\ee
That is, the system has significant bound entanglement.\footnote{In \cite{hayden2006aspects}, a regime was found in the infinite temperature case where $E_c$ is extensive while $I$ (and hence $E_d$) is sub-extensive. However, this effect took place for a range of subsystem dimensions which all correspond to $c=\ha$ in the thermodynamic limit, as opposed to an $O(1)$ range of $c$ between some $c_0$ and $\ha$, as we find in the finite temperature examples here. The conclusion about $E_c$ in \cite{hayden2006aspects} was drawn using entanglement of formation as opposed to logarithmic negativity. On calculating $\sE$ in the infinite temperature case in the same regime at $c=\ha$, we do not find that it becomes extensive while $I$ is still sub-extensive.} 

It is not clear from the above results whether the preparation of the state by PPT operations is reversible in this regime. There are two logical possibilities, both with interesting physical consequences.\footnote{The following statements also assume that $E_c^{\rm (ppt)}$ is extensive when $E_c^{\rm (ppt, exact)}$ is extensive.}
\begin{enumerate}
\item In some systems, $\sE$ also gives an estimate of $E_d^{\rm (ppt)}$ \cite{audenaert2003entanglement}. If that is the case here, then we have volume-like $E_d^{\rm (ppt)}$. This would mean that although LOCC distillable entanglement is small, PPT distillable entanglement is large, implying that finite temperature greatly enhances distillable entanglement if we use PPT operations. 
\item If $E_d^{\rm (ppt)}$ is comparable to $I$, then both the PPT and LOCC distillable entanglements are small, and the state is highly PPT-irreversible. 
\end{enumerate}

\section{Implications for an evaporating black hole}

We now turn to consequences for evaporating black holes, where we take $A$ to be the radiation and $B$ to be the remaining black hole. In~\cite{page1993information}, from averaging over random pure states, Page found that the entanglement entropy of the radiation undergoes a transition from increasing to decreasing behavior at some time scale $t= t_p$ where $ \log d_A = \log d_B$. Page's calculation was in the infinite temperature case, where from our discussion in section \ref{sec:inf_temp}, we know that $t_p$ is also the time scale at which entanglement within the radiation, as quantified by either $I$ or $\sE$, starts to become extensive. 

The natural finite-temperature generalization of the Page time is when $S_A^{(\rm eq)} = S_B^{(\rm eq)}$,\footnote{For example, this was already used in~\cite{page2013time}.} which corresponds to $c = \ha$. As discussed earlier, \eqref{page_gen} obtained from the equilibrium approximation and resolvent calculations confirms that this is indeed the time at which the entanglement entropy transitions from increasing to decreasing behaviour. However, our results for logarithmic negativity give a surprising prediction for the quantum-informational properties of the radiation at finite temperature: there are significant entanglement correlations within the radiation long before the Page time. This suggests the 
existence of another time scale $t_b$ when quantum entanglement within the radiation starts becoming extensive. 

For time scales $t \in (t_b, t_p)$, the entanglement correlations within the radiation appear to be bound entanglement,  i.e.~they cannot be distilled using LOCC. It is possible that they might be distillable using more general operations such as PPT operations. 

It is also instructive to consider the Hayden-Preskill thought experiment at a finite temperature. Suppose we throw a diary into the black hole, and see when the information of the diary is recoverable from the radiation. Applying the equilibrium approximation to the Petz recovery map, it can be shown that the information of the diary can be recovered from the radiation only after the Page time $t_p$ \cite{long_paper}. This can be viewed as giving an operational definition of the Page time. It would be desirable to also give an operational definition of the new time scale $t_b$. Such a definition may involve different quantum information tasks that can be completed using bound entanglement \cite{2006PhRvL..96o0501M}.

Note that to make predictions for the mixed-state entanglement between the black hole and some part of the radiation, we can again use the general results discussed in this paper, now taking $A_1$ and $B$ in our setup to be parts of the radiation and $A_2$ to be the black hole. 

In Euclidean gravity setups, the calculation of the negativity between parts of the radiation in the ES and ME phases would involve replica wormholes. We show these replica wormholes explicitly for the model of \cite{penington2019replica} in Appendix \ref{app:eq}. (See Fig. \ref{fig:gravity_figs}.) As discussed earlier around \eqref{jt}, due to specific features of the density of states in JT gravity, the difference between $t_b$ and $t_p$ is suppressed in the large quantity $S_0$ in this model. However, this difference should be macroscopically large for higher-dimensional black holes. 
More generally, it would be interesting to understand whether it is possible to obtain a Lorentzian derivation of the non-zero negativity before the Page time, and see whether there is any semi-classical or geometric description of bound entanglement. Another interesting question is about how new phases which might not correspond to analytic continuation, similar to the ones we found in Fig. \ref{fig:phase_diagram_a2_inf}, would be manifested in a gravity calculation.

\vspace{0.2in} \centerline{\bf{Acknowledgements}} \vspace{0.2in}
We would like to thank G. Penington for discussions. 
This work is supported by the Office of High Energy Physics of U.S. Department of Energy under grant Contract Numbers DE-SC0012567 and DE-SC0019127.

\onecolumngrid
\appendix

\begin{appendix} 
\section{Information-theoretic quantities for mixed-state entanglement}
\label{app:info}
Consider a state $\rho$ in a bipartite system $\sH_A = \sH_{A_1} \otimes \sH_{A_2}$. $\rho$ is said to be a separable state if it can be written as a convex combination of product states,
\be 
\rho = \sum_{i=1}^q p_i \, (\rho_i)_{A_1}\otimes (\tilde{\rho}_i)_{A_2} \, , \quad 0 \leq p_a \leq 1 , \quad \sum_{a=1}^q p_a = 1\, . \label{sep}
\ee
Such a state has no quantum entanglement, as the correlations in it can be given a classical hidden-variable description \cite{werner1989quantum}, and it may be prepared using only LOCC without any need for EPR pairs between $A_1$ and $A_2$. Any state $\rho$ that is not separable is said to be entangled. 

While no general criterion is known to determine whether or not an arbitrary state $\rho$ is entangled (doing so is an NP-hard problem \cite{gurvits2004classical}), we can use various quantities to study entanglement in mixed states. One familiar quantity is the mutual information
\be 
I(A_1, A_2) = S(\rho_{A_1}) +S(\rho_{A_2}) - S(\rho_{A}), \label{mut} 
\ee 
where $S(\rho)$ is the von Neumann entropy 
\be 
S(\rho) = -\text{Tr}[\rho \log \rho] 
\ee 
of $\rho$, and $\rho_{A_s}$ in \eqref{mut} refers to the reduced density matrix in subsystem $A_s$. While the mutual information is non-zero for any entangled state, it is also nonzero for the separable state \eqref{sep} with $q> 1$, and can hence reflect both classical and quantum correlations. Note that we can also define the R\'enyi mutual information in terms of the $n$-th R\'enyi entropy,
\be 
I_n(A_1, A_2) = S_n(\rho_{A_1}) +S_n(\rho_{A_2}) - S_n(\rho_{A}), \quad S_n(\rho)= -\frac{1}{n-1}\log \text{Tr}[\rho^n] \, , 
\ee 
although the physical interpretation of this quantity is not well-understood, and in particular it can take negative values.

Another useful measure is the logarithmic negativity, defined in terms of the partial transpose $\rho^{T_2}$ of $\rho$, which is given by 
\be 
\rho^{T_2}_{a_1 a_2, b_1 b_2} = \rho_{a_1 b_2, b_1 a_2}\, . 
\ee
where $a_1, b_1$ and $a_2, b_2$ are indices in $A_1$ and $A_2$ respectively. 
In terms of the eigenvalues $\lambda_i$ of $\rho_A^{T_2}$, the logarithmic negativity is defined as 
\be 
\sE(A_1, A_2) = \log \left(\sum_i |\lambda_i|\right)\, . 
\ee 
States with $\sE(A_1, A_2)> 0$ are always entangled. States with $\sE(A_1, A_2)=0$ are referred to as positive partial-transpose (PPT) states, and include the entire set of separable states, but also include some (bound) entangled states. 

From an operational perspective, two natural measures are the entanglement cost and the distillable entanglement. For both quantities, we take $n$ copies of the original system, $A_1^{\otimes n} \otimes A_2^{\otimes n}$. We allow only local operations and classical communication between $A_1^{\otimes n}$ and $A_2^{\otimes n}$, and consider conversions between $\rho^{\otimes n}$ and $(\ket{\rm EPR}\bra{\rm EPR})^{\otimes m}$, 
where 
\be 
\ket{\rm EPR} = \frac{1}{\sqrt{2}}(\ket{0}_{x_1}\ket{0}_{x_2}+ \ket{1}_{x_1}\ket{1}_{x_2}), \quad x_1, x_2 \text{ are qubits in $A_1^{\otimes n}, A_2^{\otimes n}$.} 
\ee
First consider the conversion from $(\ket{\rm EPR}\bra{\rm EPR})^{\otimes m}$ to $\rho^{\otimes n}$ under different choices $\sL$ of LOCC operations, with vanishing error in the limit $n\rightarrow \infty$. $E_c$ is defined as the minimum ratio $\frac{m}{n}$ over all choices of $\sL$ \cite{hayden2001asymptotic}. We can also require that the error in the conversion vanishes before taking the $n\rightarrow \infty$ limit, and the corresponding minimum ratio $\frac{m}{n}$ is then called the exact entanglement cost $E_c^{\rm (exact)}$ \cite{yue2019zero}. Next, consider the conversion from $\rho^{\otimes n}$ to $(\ket{\rm EPR}\bra{\rm EPR})^{\otimes m}$ under LOCC operations $\sL$. Now the maximum ratio $\frac{m}{n}$ over all choices of $\sL$ is defined as the distillable entanglement $E_d$ if we require the error to vanish only in the $n\rightarrow \infty$ limit, and the exact distillable entanglement $E_d^{\rm (exact)}$ if we require the error to vanish before taking the $n\rightarrow \infty$ limit. While for pure states $\rho_A$, 
$E_c$ and $E_d$ are both equal to the entanglement entropy $S(\rho_{A_1})$,
for mixed states in general 
$E_c \geq E_d$
and neither of these quantities must be equal to $S(\rho_{A_1})$. 

If we consider the set of PPT-preserving transformations ( i.e.~transformations that send any state $\sigma$ with $\sigma^{T_2}\geq 0$ to another state $\sigma'$ with $(\sigma')^{T_2}\geq 0$ ), of which LOCC transformations are a proper subset, then we can again consider the asymptotic rates of converting between $(\ket{\rm EPR}\bra{\rm EPR})^{\otimes m}$ and $\rho^{\otimes n}$ under such operations. The entanglement costs and distillable entanglements under such operations, $E_c^{(\rm ppt)}$, $E_d^{(\rm ppt)}$, $E_c^{(\rm ppt, exact)}$ and $E_d^{(\rm ppt, exact)}$ are then given by the natural generalizations of the definitions for LOCC in the previous paragraph. 

It is clear from the above definitions that  for a fixed set of operations, the exact costs are always greater than or equal to the costs, and the LOCC costs are always greater than or equal to the corresponding PPT costs. 

While the various entanglement costs and distillable entanglements are all difficult to compute in practice, they are related to calculable measures such as $I$ and $\sE$ through various bounds, such as \eqref{m_ed} and \eqref{e_ec}.

\section{Equilibrium approximation for R\'enyi and logarithmic negativity}
\label{app:eq}
We consider a system evolving from a far-from-equilibrium pure state $\rho_0 = \ket{\Psi_0} \bra{\Psi_0}$ to 
a state $\rho = \ket{\Psi} \bra{\Psi}$ with $\ket{\Psi} = U \ket{\Psi_0}$, which is in equilibrium at macroscopic level. We assume that the macroscopic physical properties of the equilibrated pure state $\rho$ can be approximated by an equilibrium density operator $\rho^{\rm (eq)}$ as in \eqref{heno}. 

Consider the $n^{th}$ R\'enyi entropy with respect to a subsystem $R$, 
\be \label{rnyi}
\sZ_{n, R} = e^{- (n-1) S_n^{(A)}} := {\rm Tr}_R \rho_R^n = {\rm Tr}_R \le({\rm Tr}_{\bar R} U \rho_0 U^\da \ri)^n 
= \vev{\eta_R \otimes e_{\bar R}| (U \otimes U^{\dagger})^n| \rho_0,e}
\ee
where in the last equality we have written this quantity as an amplitude in the replica space $(\sH \otimes \sH)^n$, with the following notation. 
For any operator $\sO$ acting on $\sH$, the state $\ket{\sO, \sig} \in (\sH \otimes \sH)^n$, where $\sig$ is an element of the permutation group $\sS_n$ of $n$ objects, is defined as 
\bega \label{aa1}
\vev{i_1 \bar i_1' i_2 \bar i_2' \cdots i_n \bar i_n'| \sO, \sig} = \sO_{i_1 i'_{\sig (1)}} \sO_{i_2 i'_{\sig (2)}} \cdots 
\sO_{i_n i'_{\sig (n)}} , \quad \sO_{ij} = \vev{i |\sO|j} \ .
\end{gather} 
Here $\{\ket{i_1 \bar i_1' i_2 \bar i_2' \cdots i_n \bar i_n'}\}$ is a 
basis for $(\sH \otimes \sH)^{n}$, and $\sig (i)$ denotes the image of $i$ under $\sig$. 
For $\sO$ given by the identity operator, we will denote the states obtained in this way simply as $\ket{\sig}$. 
When the system is divided into subsystems, we can similarly define states by associating different permutations to different subsystems. For example, suppose $\sH = \sH_A \otimes \sH_{\bar A}$. Then $\ket{\sO, \tau_A \otimes \sig_{\bar A}}$ with $\tau, \sig \in \sS_n$ is defined as 
\bega 
 \vev{i_{1_a} i_{1_b} \bar i_{1_a}' \bar i_{1_b}' \cdots i_{n_a} i_{n_b} \bar i_{n_a}' \bar i_{n_b}' |\sO, \tau_A \otimes \sig_{\bar A}} = 
 \sO_{i_{1_a} i_{1_b}, i_{\tau(1)_a}' i_{\sig(1)_b}'} \cdots \sO_{i_{n_a} i_{n_b}, i_{\tau(n)_a}' i_{\sig(n)_b}'} 
\label{zn_t}
\end{gather}
where $\ket{i_{k_a}}, \ket{\bar i'_{k_a}}$, $\ket{i_{k_b}}, \ket{\bar i'_{k_b}}$ respectively denote basis vectors for subsystems $A$ and $\bar A$ in the $k^{th}$ replica of $\sH \otimes \sH$.
In~\eqref{rnyi}, for the state $\ket{\eta_R \otimes e_{\bar R}}$, $\sO$ is the identity operator $\mathbf{1}$, $e$ identity permutation and $\eta$ the cyclic permutation $(n, n-1, \cdots 1)$, which sends $n$ to $n-1$, $n-1$ to $n-2$ and so on. 

As explained in \cite{liu2021entanglement}, we can find the approximate late-time value of $\sZ_{n, R}$ in a chaotic system by inserting a physically motivated projection in \eqref{rnyi}. The projection involves the effective identity operator $\sI_{\al}$ associated with the macroscopic equilibrium of $\rho$, and leads to the expression \eqref{fen}. 

\begin{figure}[!h]
 \centering
 \includegraphics[width=10cm]{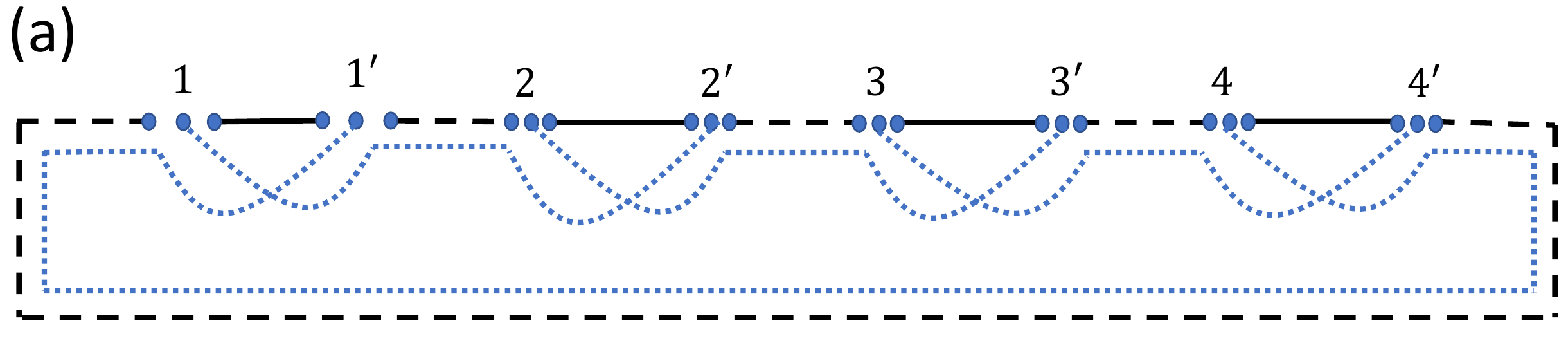}
 \includegraphics[width=10cm]{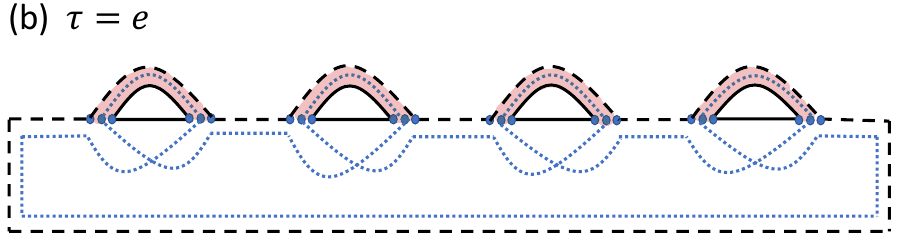}
 \includegraphics[width=10cm]{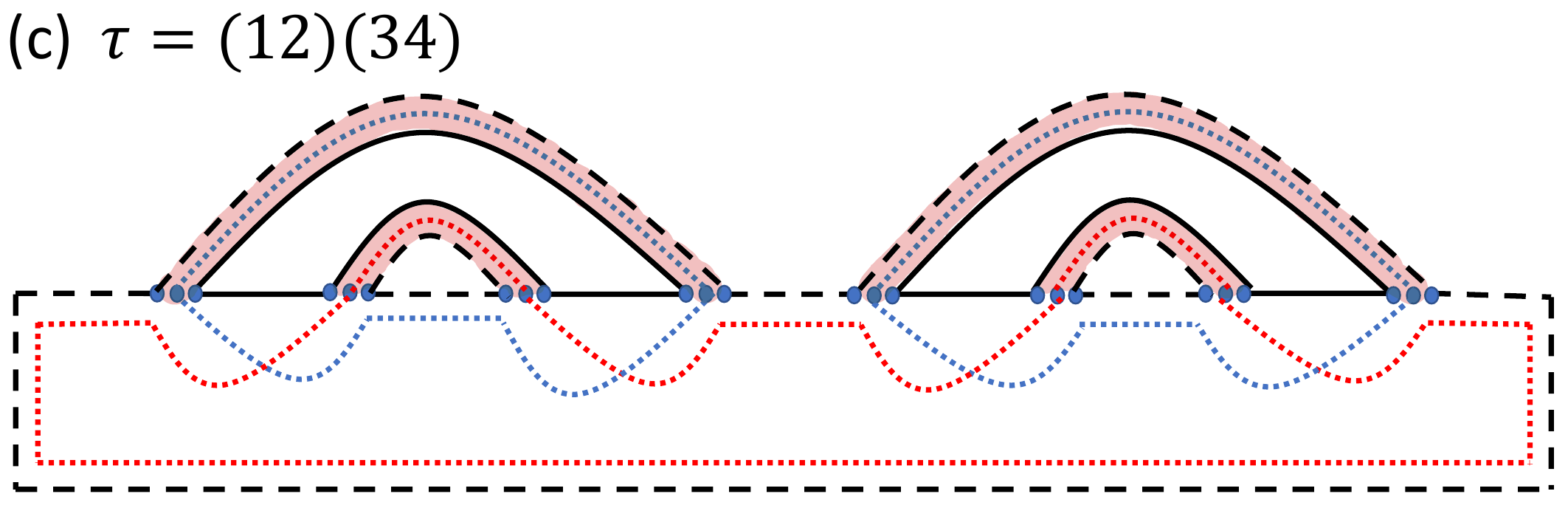}
 \includegraphics[width=10cm]{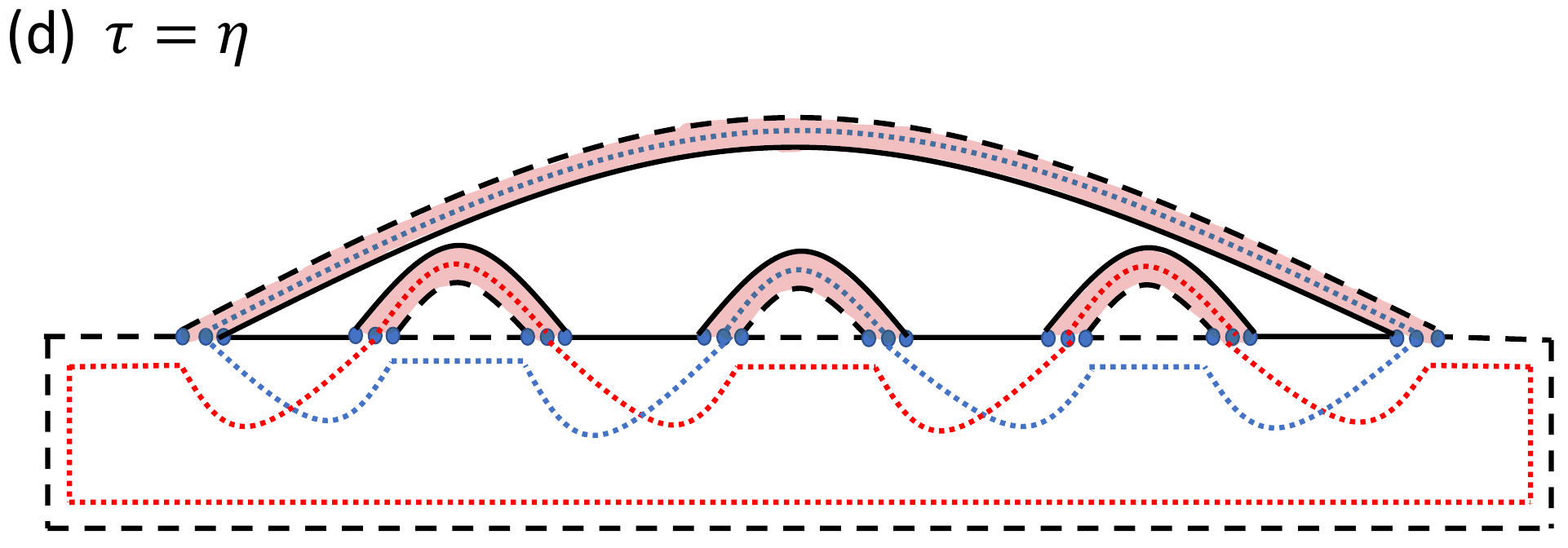}
 \caption{(a) shows the common lines of all diagrams for different terms in \eqref{repa}, and (b)-(d) show examples for three choices of $\tau$ which are dominant in various regimes, for the case $n=4$. The dotted loops are shown in two different colors in (c) to make the distinction between different loops clear.}
 \label{fig:negativity_figs}
\end{figure}

For the partial transpose partition function $\sZ_n^{\rm (PT)}$, a similar set of steps starting from an expression for $\sZ_n^{\rm (PT)}$ as an amplitude in $(\sH \otimes \sH)^n$ leads to the expression \eqref{repa}, where $\eta^{-1}$ is permutation $(1 \, 2\, ... \, n-1 \, n)$. Each term in \eqref{repa} can be given a diagrammatic representation, as shown in Fig.~\ref{fig:negativity_figs}. These diagrams make it easier to understand the dependence of different terms on $\sI_{\al}$ and the sizes of various subsystems. We can insert the identity in the terms of \eqref{fen} to write 
\be 
\label{negativity_diag}
\begin{gathered} 
\vev{\eta_{A_1} \otimes \eta^{-1}_{A_2} \otimes e_{B} | \sI_\al , \tau} = \sum_{i_1, i'_1, ... i_n, i'_n} \vev{\eta_{A_1} \otimes \eta^{-1}_{A_2} \otimes e_{B}|i_1 \bar{i'}_1 ... i_n \bar{i'}_n} \vev{i_1 \bar{i'}_1 ... i_n \bar{i'}_n| \sI_{\al}, \tau} \, , \\
\ket{i_m} = \ket{i_{m_a}}_{A_1} \ket{i_{m_{\bar{a}}}}_{A_2} \ket{i_{m_b}}_B, \quad \ket{\bar{i'}_m} = \ket{\bar{i'}_{m_a}}_{A_1} \ket{\bar{i'}_{m_{\bar{a}}}}_{A_2} \ket{\bar{i'}_{m_b}}_B \,.
\end{gathered} 
\ee
 The lower half of each diagram represents $\vev{\eta_{A_1} \otimes \eta^{-1}_{A_2} \otimes e_{B}|i_1 \bar{i'}_1 ... i_n \bar{i'}_n}$ by connecting $i_{m_a}$ with $i'_{\eta(m)_a}$ using dashed lines, $i_{m_{\bar{a}}}$ with $i'_{\eta^{-1}(m)_{\bar{a}}}$ using dotted lines, and $i_{m_b}$ with $i'_{m_b}$ using solid lines, as shown in Fig.~\ref{fig:negativity_figs}(a). The upper half of the diagram represents 
$\vev{i_1 \bar{i'}_1 ... i_n \bar{i'}_n| \sI_{\al}, \tau}$, by connecting $i_m$ with $i'_{\tau(m)}$, as shown for some examples in Fig.~\ref{fig:negativity_figs}. There is a similar diagrammatic representation for \eqref{fen} which was explained in \cite{liu2021entanglement}. 

In the thermodynamic limit, the RHS of \eqref{fen} and \eqref{repa} can be approximated by the terms from some subset of permutations that give the dominant contribution. For the R\'enyi partition function \eqref{fen}, the dominant contribution is always given by either $\tau=e$ or $\tau= \eta$, leading to 
\be 
S_{n}(\rho_R) = \text{min}(S_{n, R}^{\rm (eq)}, S_{n, \bar{R}}^{\rm (eq)}) 
\ee
where $S_{n, R}^{\rm (eq)}$ is the $n^{th}$ R\'enyi entropy of $R$ for the state $\rho^{\rm (eq)}$. For the partial transpose partition function $\sZ_n^{\rm (PT)}$ with even $n$, the dominant contribution for all choices of $\sI_{\al}$ is given in different parts of the phase diagram by one out of $\tau= e$, $\tau= \eta \text{ or }\eta^{-1}$, and $\tau= \tau_{ES}$, where $\tau_{ES}$ refers to the two permutations with two-cycles among adjacent elements, 
\be 
\tau_{ES}=\{(12)(34)...(n-1\, n),~ (23)(45)...(n\, 1) \}
\ee
The diagrams corresponding to $e$, $\eta$ and one of the permutations in $\tau_{ES}$ for $n=4$ are shown in Fig. \ref{fig:negativity_figs}. 
On analytic continuation to $n\rightarrow 1$, the contributions to $\sZ_n^{\rm (PT)}$ from these permutations respectively give the finite-temperature expressions for the negativity in the NE, ME and ES phases in \eqref{fin_zero}-\eqref{fin_es}, \eqref{fin2_zero}-\eqref{fin2_es}, and \eqref{fin3_zero}-\eqref{fin3_es}. 

\begin{figure}[!h]
 \centering
 \includegraphics[width=12cm]{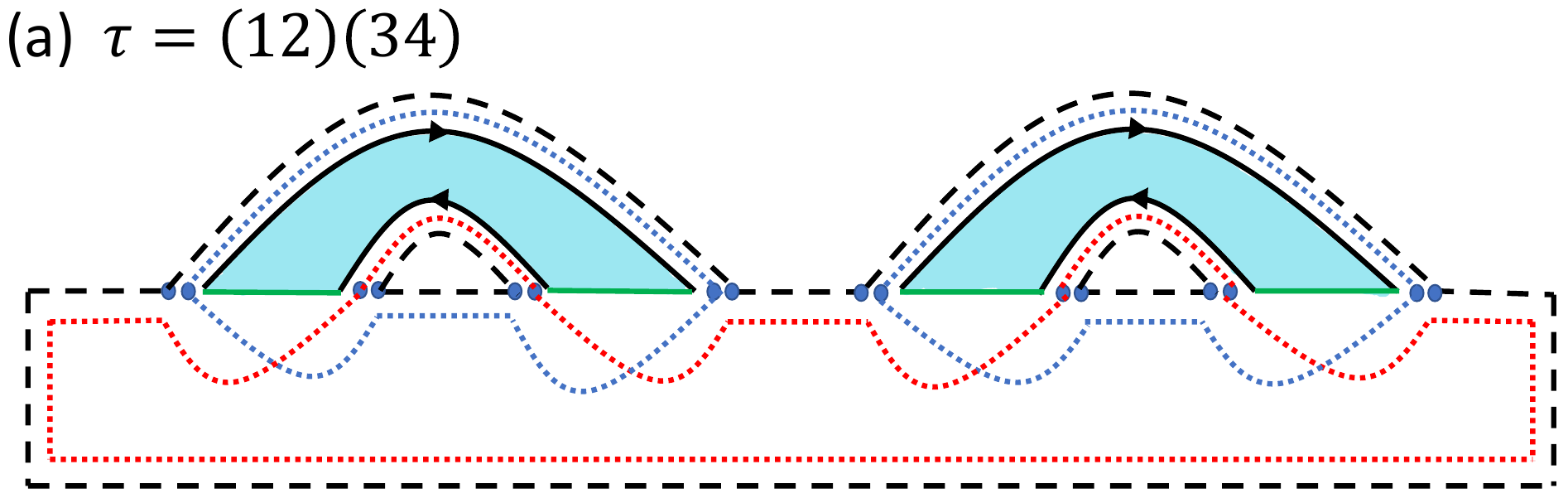}
 \includegraphics[width=12cm]{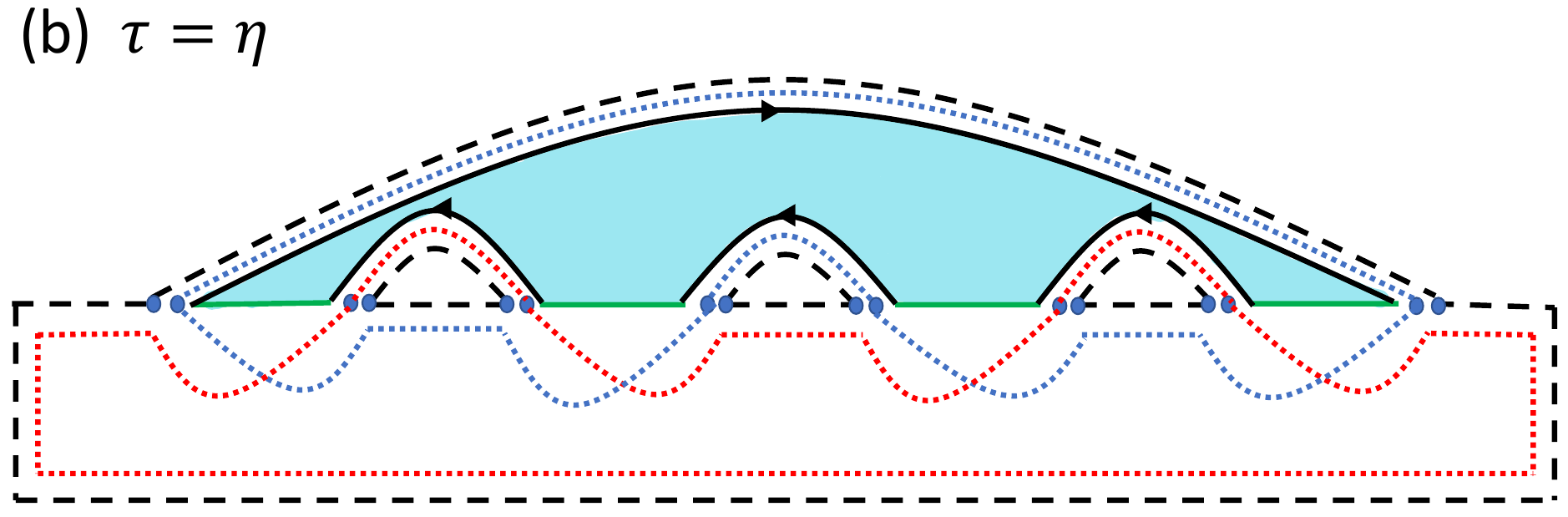}
 \caption{Euclidean gravity path integrals with replica wormholes for calculating $\sZ_n^{\rm (PT)}$ in the model of \cite{penington2019replica}, for $n=4$. The dominant contribution in the ES phase, shown in (a), involves connected partition functions $Z_{2, B}$ between two asymptotic boundaries, while the ME phase contribution in (b) involves $Z_{n, B}$. The black lines with arrows are asymptotic boundaries in JT gravity (each with length $\beta$, although the lengths appear to differ in the figure), and the green lines are EOW branes. The dashed and dotted loops simply give factors of $d_{A_1}$ and $d_{A_2}$ respectively.}
 \label{fig:gravity_figs}
\end{figure}

In gravity setups, where we take $B$ to be the black hole subsystem and $A$ to be the radiation, the contributions to $\sZ_n^{\rm (PT)}$ from the permutations $\tau_{ES}$, $\eta$, and $\eta^{-1}$ all involve replica wormholes. For example, consider the model for black hole evaporation in \cite{penington2019replica} where the black hole subsystem consists of JT gravity with EOW (end-of-the-world) branes. The boundary calculation of $\sZ_n^{\rm (PT)}$ between two parts of the radiation in this model is precisely equivalent to \eqref{repa}, with effective identity operator
\be 
\sI_{\al} = \mathbf{1}_A \otimes e^{-\beta H_B} f(H_B)\, . 
\ee
This is a special case of \eqref{can}, with $A$ at infinite temperature and an additional factor $f(H_B)$ which captures contributions from end-of-the-world (EOW) branes. In this example, we can read off the boundary expressions for $\sZ_n^{\rm (PT)}(\tau)$ in terms of $d_{A_1}, d_{A_2}, Z_{n, B}$ from the diagrams in Fig. \ref{fig:negativity_figs} (where $Z_{n, B}= \text{Tr}[(e^{-\beta H_B} f(H_B))^n]$). Using the relation between bulk and boundary partition functions in holography, we then find geometries like the ones shown in Fig. \ref{fig:gravity_figs} for evaluating $\sZ_n^{\rm (PT)}$ in the ES and ME phases. Both involve connected Euclidean gravity path integrals between multiple asymptotic boundaries. More generally, \eqref{repa} can be used to systematically derive replica wormhole prescriptions for calculating $\sZ_n^{\rm (PT)}$ in other gravity setups. 

\section{Calculation of logarithmic negativity through the resolvent}
\label{app:resolvent}

In order to find the von Neumann entropy and the logarithmic negativity, especially in the cases at finite temperature where we cannot a priori use analytic continuation, it is useful to first find the equilibrium approximation for the resolvents in \eqref{Rdef} and \eqref{Rndef}. Below we will explicitly discuss how to find $R_N$; the calculation for the resolvent for the von Neumann entropy is similar. 

\begin{figure}[!h] 
\centering
\includegraphics[width=12cm]{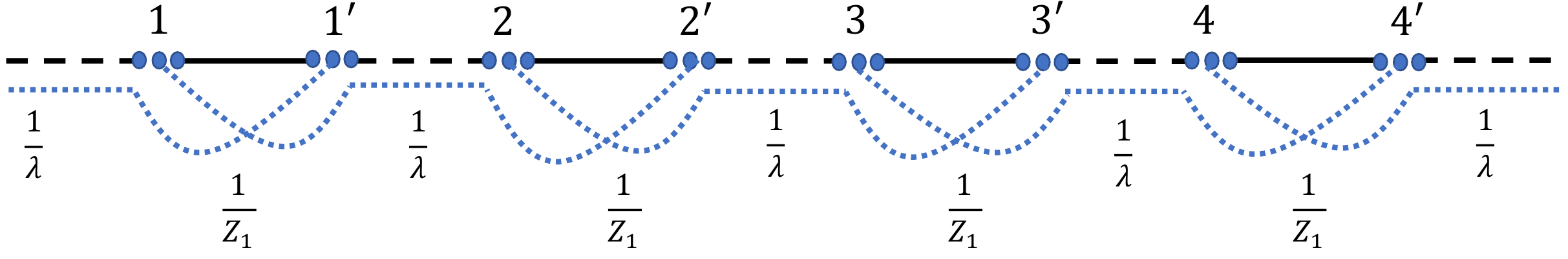}
\caption{``Boundary" lines for the equilibrium approximation for $R_{pq}$ with $n=4$.}
\label{fig:rij_ext}
\end{figure}

\begin{figure}[!h] 
\centering
\includegraphics[width=12cm]{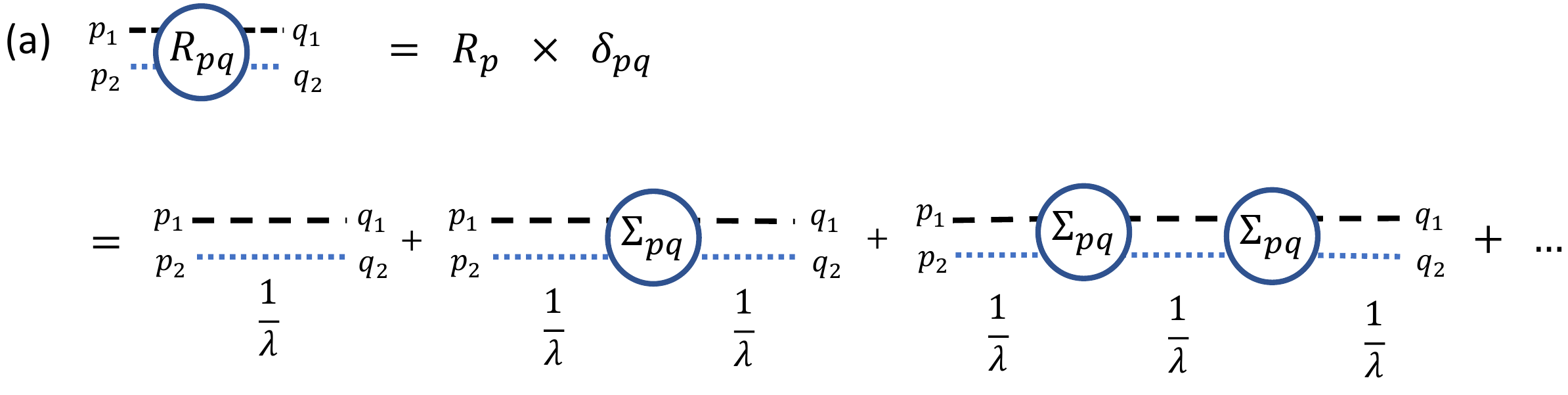}
\includegraphics[width=12cm]{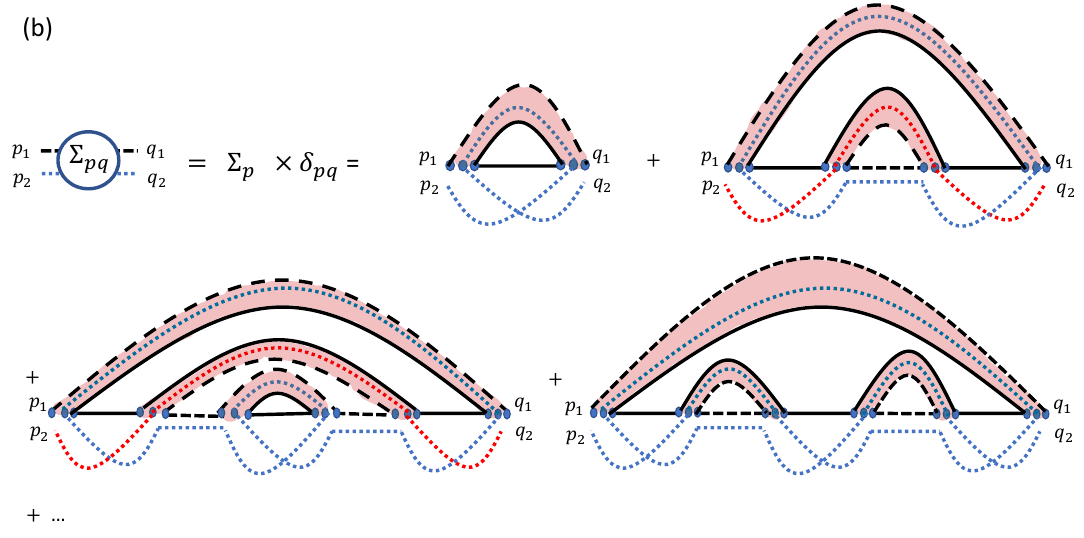}
\caption{Diagrammatic representation of $R_{pq}$ and the first few diagrams contributing to $\Sigma_{pq}$ in the case where $A_1$ is larger than $A_2$. }
\label{fig:rpq}
\end{figure}

It is useful to see $R_N$ as the trace of a matrix 
\be 
R_{p q} =  \frac{1}{\lambda}\sum_{n=0}^{\infty} \frac{1}{\lambda^n} {(\rho_A^{T_2})^n}_{pq} , \quad \ket{p} = \ket{p_1}_{A_1} \ket{p_2}_{A_2} , \quad \ket{q} = \ket{q_1}_{A_1} \ket{q_2}_{A_2} \label{rpq_def}
\ee
We can apply the equilibrium approximation to each ${(\rho_A^{T_2})^n}_{pq}$. The common lower half of the diagrams for all permutations in this case can be deduced from Fig. \ref{fig:negativity_figs} (a) for $\text{Tr}[(\rho_A^{T_2})^n]$ by erasing the dashed line connecting $i'_{a_{n}}$ and $i_{a_1}$, and the dotted line connecting $i'_{\bar{a}_{1}}$ and $i_{\bar{a}_n}$, and instead taking the inner product of $\ket{i_{a_1}}$, $\ket{i'_{\bar{a}_{1}}}$, $\ket{i_{a_n}}$, $\ket{i'_{\bar{a}_{n}}}$ with $\ket{p_1}$, $\ket{p_2}$, $\ket{q_1}$, $\ket{q_2}$ respectively. The resulting lines are shown in Fig. \ref{fig:rij_ext}, which also explains how the factors of $\frac{1}{\lambda^{n+1}}$ and $\frac{1}{Z_1^n}$ that are common to all terms for a given $n$ in \eqref{rpq_def} (the second factor comes from the equilibrium approximation) are incorporated into these lines. 
In the limit where the effective dimension of $A_1$ is much larger than that of $A_2$, it is sufficient to consider contributions from planar diagrams for all $n$ to $R_{pq}$. We can write $R_{pq}$ in terms of a self-energy $\Sigma_{pq}$ as shown in Fig. \ref{fig:rpq}(a). $\Sigma_{pq}$ is a sum of diagrams without any disconnected parts connected by $\frac{1}{\lambda}\delta_{pq}$, to which the first few contributions are shown in Fig.~\ref{fig:rpq}(b). We take $\ket{p_1}, \ket{q_1}$ and $\ket{p_2}, \ket{q_2}$ to be elements of the energy eigenbasis in $A_1$ and $A_2$ respectively, so that we approximately have that 
\be 
 \bra{p_1}\bra{p_2}\sI_{\al} \ket{q_1} \ket{q_2} \propto \delta_{p_1 \, q_1}\, \delta_{p_2 \, q_2}
\ee
for both the canonical and microcanonical ensembles, and hence from the diagrams contributing to $\Sigma_{pq}$ we can see that both $\Sigma_{pq}$ and $R_{pq}$ are proportional to $\delta_{pq}$. 

\begin{figure}[!h] 
\centering
\includegraphics[width=15cm]{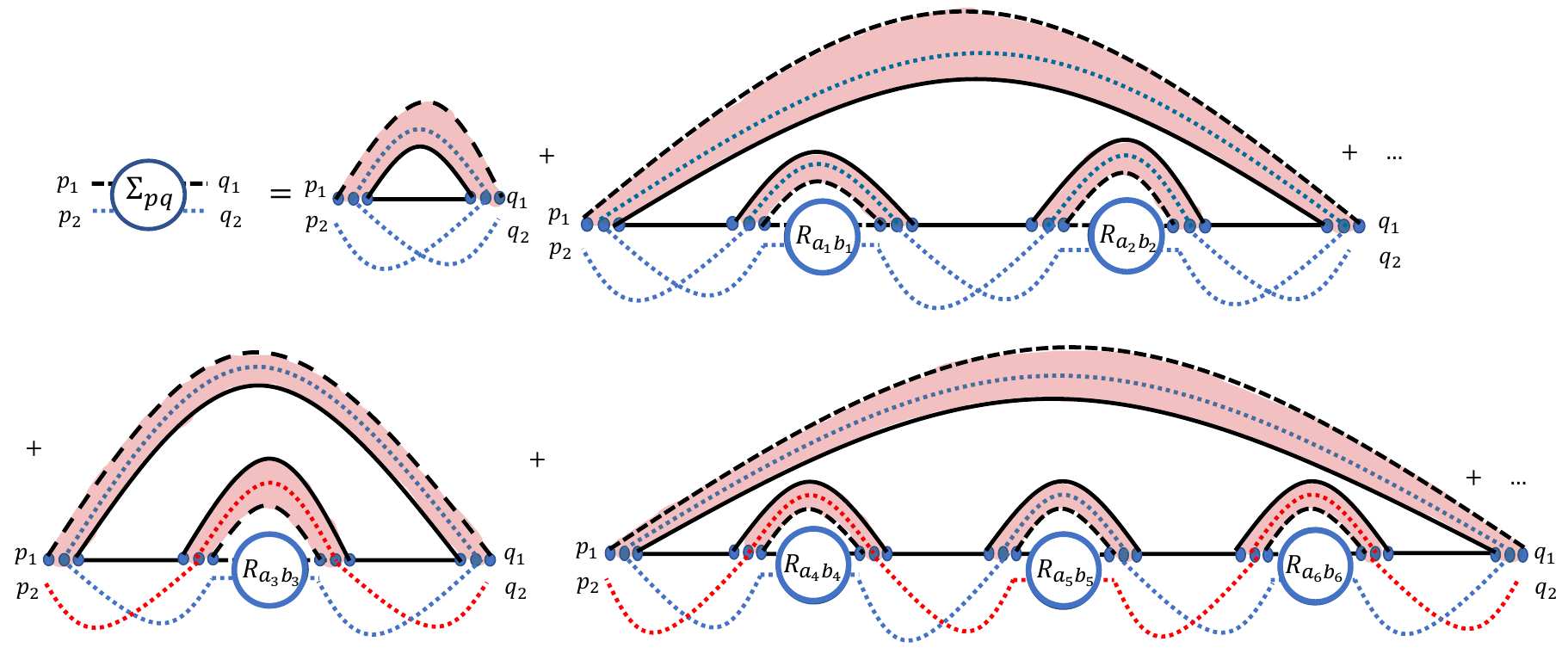}
\caption{Rewriting of the RHS of Fig. \ref{fig:rpq}(b) as a Schwinger-Dyson equation. All $a_i, b_i$ are independently summed over (unless they have delta functions among each other or with $p, q$ according to the diagram).}
\label{fig:rpq_sd}
\end{figure}

We can immediately see by summing the geometric series on the RHS of Fig. \ref{fig:rpq}(a) that 
\be 
R_N = \sum_p R_p , \quad R_{p} = \frac{1}{\lambda - \Sigma_{p}} \, . 
\ee
We can systematically include all planar contributions to $\Sigma_{p}$ using the Schwinger-Dyson equation shown diagrammatically in Fig. \ref{fig:rpq_sd}. For general choices of $\sI_{\al}$, this Schwinger-Dyson equation in general leads to a complicated set of equations relating $R_p$ for all different $p$ to each other. Below we consider a few choices of $\sI_{\al}$ for which the Schwinger-Dyson equation simplifies. More details of each of these calculations will be discussed in \cite{long_paper}. 

\subsection{Infinite temperature}
\label{sec:inf}
Taking $\sI_{\al}$ to be the identity operator on the full system, the Schwinger-Dyson equation for $R_p$ becomes independent of the index $p$ in $A$, and we have 
\be 
R_N = d_A R_1, \quad R_1 = \frac{1}{\lambda - \Sigma_1}\, . \label{r1_simple}
\ee
Each line on the RHS of Fig. \ref{fig:rpq_sd} now simplifies to a geometric series, and we get a cubic equation for $R_1$, 
\be 
\beta \lambda R_1^3 + (\alpha-\beta) R_1^2 + \left(\frac{1}{d_A}-\lambda\right) R_1+ 1= 0, \quad \alpha = \frac{1}{d_A d_B}, \quad \beta = \frac{1}{(d_{A_2}d_B)^2}
 \, . \label{rinf_cubic}
\ee
$D_N(\lambda)$ and $\sE$ can be found numerically from the solution to \eqref{rinf_cubic}, and turn out to agree with the analytic continuation in \eqref{ehn}, \eqref{legn}, \eqref{neg_inf_es}, as discussed in \cite{2021PRXQ....2c0347S}. 

Now consider the regime where $d_{A_1}/(d_{A_2} d_B)\ll 1$. This corresponds to being outside the ME phase. Then since in this regime $\alpha \gg \beta$, $R_1$ is $O(1)$, and $\lambda$ is $O(1/d_A)$, \eqref{rinf_cubic} simplifies to a quadratic equation for $R_1$, 
\be 
 \alpha \, R_1^2 + \left(\frac{1}{d_A}-\lambda\right) R_1+ 1= 0 . \label{r1_quad}
\ee
The same quadratic equation can be obtained diagrammatically by ignoring all contributions to the Schwinger-Dyson equation in Fig. \ref{fig:rpq_sd} except the first term in each line on the RHS. This corresponds to including contributions to $\sZ_n^{\rm (PT)}$ for all $n$ from permutations where all cycles have either one or two elements (including $\tau=e$ and $\tau= \tau_{ES}$, as well as other permutations such as $(12)$). We can now solve \eqref{r1_quad} to get a simple semicircle form of $D_N(\lambda)$ from $R_N(\lambda)$, 
 which can be integrated analytically to get the NE phase \eqref{ehn} and the ES phase \eqref{neg_inf_es}, and the correct transition line between them at $c= \ha$. But as expected, this approximation misses the ME phase \eqref{legn}.

\subsection{Canonical ensemble with infinite temperature in $A$}
\begin{figure}[!h] 
\centering 
\includegraphics[width=.48\textwidth]{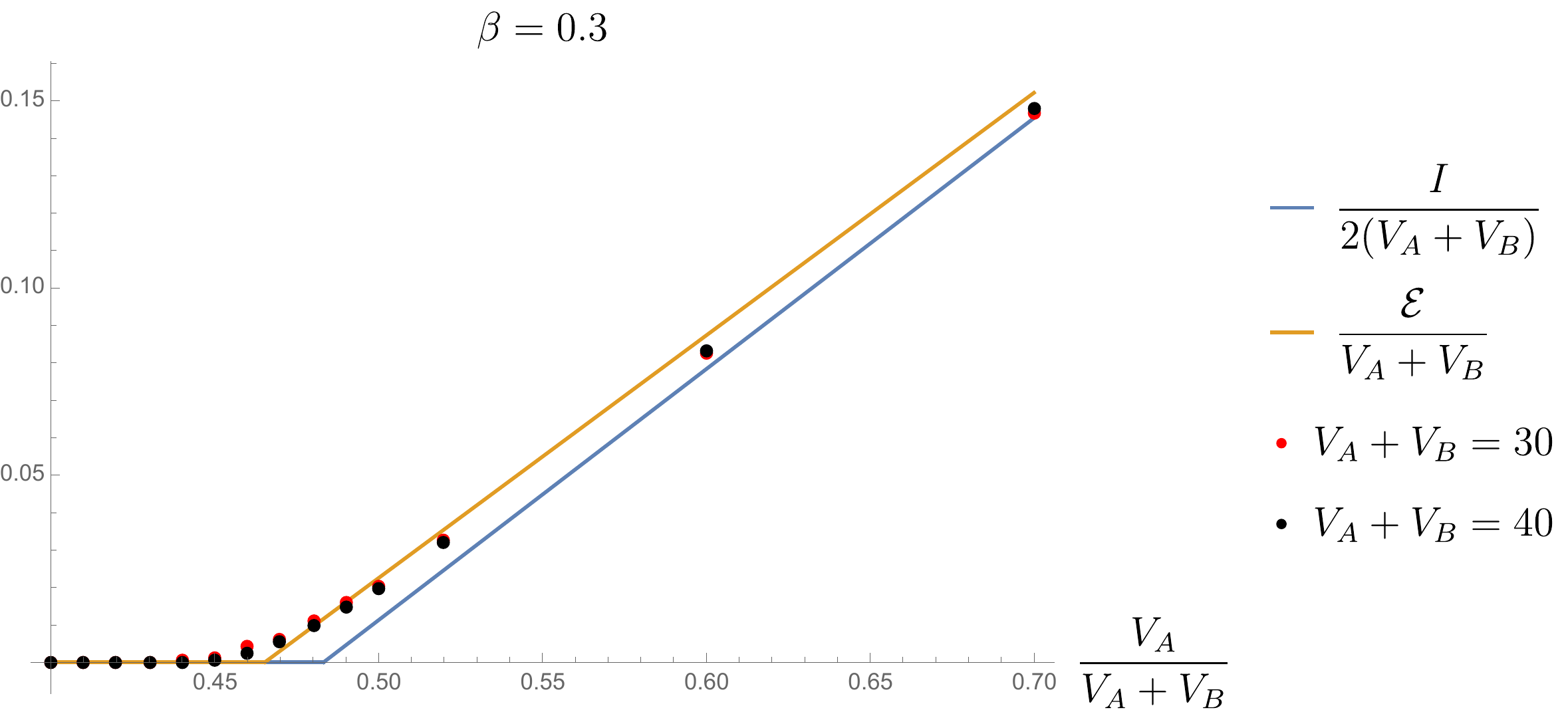}
\includegraphics[width=.48\textwidth]{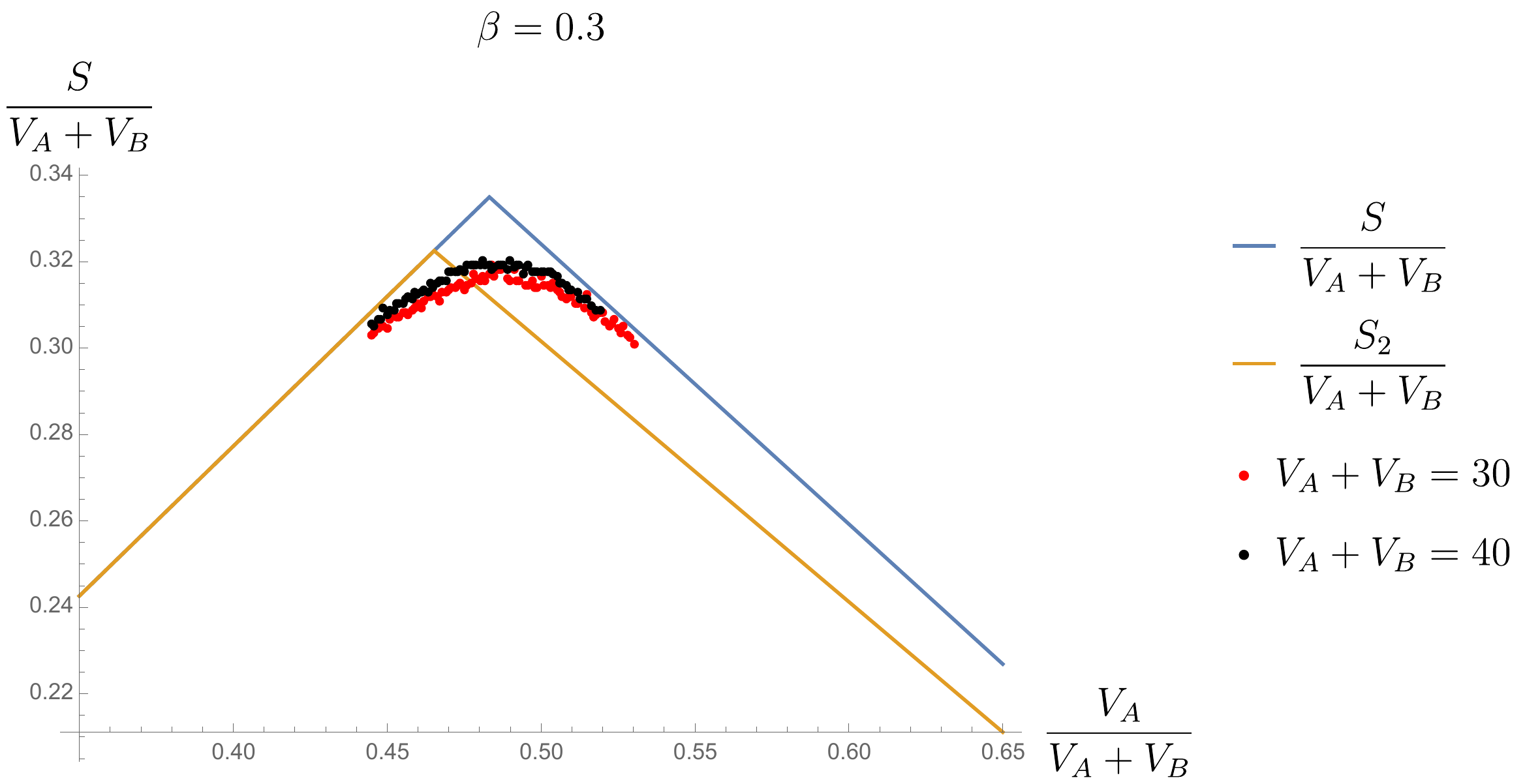}
\caption{Left: For \eqref{can} with $A$ at infinite temperature, the logarithmic negativity is computed numerically using the resolvent \eqref{R_negativity} and compared to the naive analytic continuation in \eqref{fin_zero} and \eqref{fin_es}, with excellent agreement. Right: The same is done for the von Neumann entropy of $A_1 = A$ using the resolvent \eqref{R_entropy}. Clearly, the Page transition occurs at the same place as the naive analytic continuation in \eqref{page_gen} and not at, for instance, the transition for the second R\'enyi entropy. For simplicity, we have used a Gaussian density of states $\rho(E) \propto \exp\left[-\frac{E^2}{2N_B}\right]$ and taken $\lambda = \ha$. 
}
\label{fig:resolvent_checks}
\end{figure}

Next, consider $\sI_{\al}$ as in \eqref{can}, with infinite temperature in $A$. For this case, the RHS of Fig. \ref{fig:rpq_sd} implies that $R_p$, $\Sigma_p$ again become independent of the index $p$, so that we have \eqref{r1_simple} again, but now with $\Sigma_1$ dependent on the partition functions $Z_{n, B} = \text{Tr}[e^{-n \beta H_B }]$. We find for this case 
\begin{align}
  \lambda R_N = d_A + d_{A_2}\int dE \rho(E) \sum_{k = 1}^{\infty} \left[\left(\frac{R_Ne^{-\beta E}}{d_A d_{A_2}Z_{1,B}} \right)^{2k-1}+ d_{A_2} \left(\frac{R_Ne^{-\beta E}}{d_A d_{A_2}Z_{1,B}} \right)^{2k}\right].
\end{align}
We then complete the geometric sums to find
\begin{align}
  \lambda R_N = d_A + \int dE \rho(E)\frac{d_{A_2}^2 R_N \left(d_{A} Z_{1,B} e^{\beta E}+R_N\right)}{d_{A}^2
  d_{A_2}^2 Z_{1,B}^2 e^{2 \beta E}-R_N^2}.
  \label{R_negativity}
\end{align}
where $\rho(E) = e^{V s(E/V)}$ is the density of states for $B$. On specifying the density of states, this equation can be can be solved numerically for $R_N(\lambda)$ and used to obtain $\sE$. The solution with a gaussian entropy density is shown in Fig. \ref{fig:resolvent_checks}~(a) as a function of $c$ at $\lambda = \ha$. The result agrees with the expressions for $\sE$ in the ME and ES phases in \eqref{fin_zero} and \eqref{fin_es} and the naive phase transition line between them in \eqref{ehv}. 

Similar to the infinite temperature case, we can obtain an analytic expression for the regime where the effective dimensions $d_{A_1}, d_{A_2}, d_{B}$ of the different subsystems are such that $\frac{d_{A_1}}{d_{A_2} d_B}\ll 1$. Assuming that the same set of permutations contributes in this regime at finite temperature as the one that contributed at infinite temperature, we can then include only the first diagram from each line on the RHS of Fig. \ref{fig:rpq_sd}. We again find a semicircle form of $D_N(\lambda)$, which can be integrated analytically to get \eqref{fin_zero} and \eqref{fin_es} and the transition \eqref{ehv}. But note that in this case we do not have a systematic way of justifying the truncation to the permutations with cycles of only length one and two, so it is important to confirm these expressions numerically using the full resolvent as in Fig.~\ref{fig:resolvent_checks} (a).  

Analogously to the above steps, we may find an integral expression for the resolvent for the von Neumann entropy of $\rho_{A_1}$
\begin{align}
  \lambda \sR &= d_{A_1} + \int dE \rho (E)\sum_{k = 1}^{\infty}\left( \frac{\sR e^{-\beta E}}{d_{A}Z_{1,B}}\right)^k
  \quad = d_{A_1} + \int dE \rho (E)\frac{\sR}{d_A Z_{1,B} e^{\beta E}-\sR}.
  \label{R_entropy}
\end{align}
The von Neumann entropy can again be evaluated numerically, confirming the expectation from analytic continuation in \eqref{page_gen}, as shown in Fig. \ref{fig:resolvent_checks} (b).

\subsection{Microcanonical ensemble with $B$ at infinite temperature}
In this case, $R_p$ is no longer independent of $p$, but instead depends on the energies of $\ket{p_1}, \ket{p_2}$. The resolvent calculation for this case becomes more complicated, and we are not able to find $\sE$ from the sum over all permutations either analytically or numerically. But if we again truncate to the contributions from permutations that have cycles with only one and two elements, assuming this truncation is valid away from the ME phase, then this results in a form of $D_N(\lambda)$ which can be integrated analytically to confirm \eqref{fin2_zero} and \eqref{fin2_es}. The naive transition line where we set \eqref{fin2_es} to zero is also confirmed by this calculation. 

\subsection{Microcanonical ensemble with $A_2$ at infinite temperature}
In this case, the expressions for $\sZ_n^{\rm (PT)}$ are such that the resolvent can be expressed in a simple way in terms of the infinite-temperature resolvent from \eqref{rinf_cubic}. As a result, the quantity $\sZ^{\rm (PT)} \equiv \text{exp}\, {\sE}$ can also be expressed in a simple way in terms of its infinite temperature value $\sZ^{\rm (PT)}_{\infty}(d_{A_1}, d_{A_2}, d_B)$, 
\be 
\begin{gathered} 
\sZ^{\rm (PT)} \approx \sum_{E_1} p_{E_1}\, \sZ^{\rm (PT)}_{\infty}(d^{A_1}_{E_1},\, d_{A_2}, \, d^B_{E-E_1})
\label{4_cases}
\end{gathered} 
\ee 
where $d^R_{E_R} \equiv e^{V_R \, s(E_R/ V_R)}$ refers to the density of states in subsystem $R$ at energy $E_R$, with $s(\epsilon)$ the entropy density for the system, and 
\be 
p_{E_1} = \frac{d^{A_1}_{E_1} d^B_{E-E_1}}{N_E},\quad N_E = \sum_{E'_1} d^{A_1}_{E'_1} d^B_{E-E'_1} \,. 
\ee
 In the thermodynamic limit, for certain ranges of $c$ and $\lambda$, \eqref{4_cases} gives the expressions expected from analytic continuation in \eqref{fin3_zero}-\eqref{fin3_es}. In addition to these, we get two new phases, where 
\begin{align} 
\sE_{ES-ME1} = \log d_{A_2} + V_{A_1} (s(\theta_3)- s(\epsilon)) - V_B s(\epsilon) \,. 
\label{ln_4_mt}\\ 
\sE_{ES-ME2} = V_{A_1} \, (2 s(\theta_2)- s(\epsilon)) - V_B \, s(\epsilon) \, . 
\label{ln_5_mt}
\end{align} 
with $\epsilon = \frac{E}{V_{A_1}+V_B}$, and $\theta_2$ and $\theta_3$ defined implicitly as solutions to the equations 
\be 
\begin{gathered} 
\log d_{A_2} + V_B\, s\left(\frac{\epsilon\, (V_{A_1}+ V_B)- V_{A_1}\, \theta_2}{V_{B}} \right) = V_{A_1} s(\theta_2), \\
V_{A_1}\, s(\theta_3) + V_B\, s\left(\frac{\epsilon\, (V_{A_1}+ V_B)- V_{A_1}\, \theta_3}{V_{B}} \right) = \log d_2\, . 
\end{gathered} 
\ee
The full phase diagram is shown in Fig.~\ref{fig:phase_diagram_a2_inf}.

\end{appendix}


\newpage

\end{document}